\documentclass[twocolumn,nofootinbib,amsmath,amssymb,aps,prd,balancelastpage,superscriptaddress]{revtex4-1}

\usepackage{color}
\usepackage{xcolor}
\usepackage[active]{srcltx}
\usepackage{amsmath,amsfonts,amssymb,amsthm,amstext,amscd,eucal,srcltx}
\usepackage{epsfig,graphicx,bm}
\usepackage{epstopdf, epsf}
\usepackage{dcolumn}
\usepackage{comment}
\usepackage{hyperref}
\usepackage{tikz}
\usepackage[nointegrals]{wasysym}
\usepackage{braket}
\usepackage{etoolbox}
\DeclareMathOperator{\tr}{tr}

\newcommand{\be}{\begin{equation}}
\newcommand{\ee}{\end{equation}}

\newcommand{\bse}{\begin{subequations}}
\newcommand{\ese}{\end{subequations}}
\newcommand{\ba}{\begin{eqnarray}}
\newcommand{\ea}{\end{eqnarray}}
\newcommand{\bc}{\begin{center}}
\newcommand{\ec}{\end{center}}

\newcommand{\nn}{\nonumber}
\newcommand{\e}{\mathrm{e}}

\def\bra#1{\langle#1|}
\def\ket#1{|#1\rangle}
\def\braket#1{\langle#1\rangle}

\begin{document}
\preprint{IPM/P-2012/009}  
\vspace*{3mm}

\title{Scrambling Power of Soft Photons}%

\author{Xuan-Lin Su}
\email{19110190015@fudan.edu.cn}
\affiliation{Center for Field Theory and Particle Physics \& Department of Physics, Fudan University, 200433 Shanghai, China}

\author{Alioscia Hamma}
\email{alioscia.hamma@unina.it}
\affiliation{Dipartimento di Fisica Ettore Pancini, Universit\`a degli Studi di Napoli Federico II,
Via Cintia, 80126 Napoli NA}
\affiliation{INFN, Sezione di Napoli, Italy}

\author{Antonino Marcian\`o}
\email{marciano@fudan.edu.cn}
\affiliation{Center for Field Theory and Particle Physics \& Department of Physics, Fudan University, 200433 Shanghai, China}
\affiliation{Laboratori Nazionali di Frascati INFN, Frascati (Rome), Italy, EU}

\begin{abstract}
\noindent
Observable scattering processes entail emission-absorption of soft photons. As these degrees of freedom go undetected, some information is lost. 
Whether some of this information can be recovered in the observation of the hard photons, depends of the actual pattern of the scrambling of information. We compute the information scrambling of photon scattering by the tripartite mutual information in terms of the $2$-Renyi entropy, 
and find a finite amount of scrambling is present. The developed procedure thus sheds novel light on the black hole information loss paradox, showing that scrambling is a byproduct of decoherence achieved by the scattering system in its interaction with the environment, due to the emission-absorption of soft photons in fully unitary processes.
\end{abstract}

\maketitle

\section{Introduction}\label{intro}
\noindent
The infrared catastrophe is an old-standing open problem in quantum field theory (QFT) \cite{QFTMS}. This is related to the integration of massless virtual bosons in the $S$ matrix, which causes the $S$ matrix itself to diverge. There are two main methods to deal with infrared divergences, namely  the inclusive probability formalism and the dressed state formalism. The inclusive probability formalism was originally proposed for electron photon scattering process in quantum electrodynamics (QED) by Bloch and Nordsieck \cite{PR5254}, then delved by Yennie, Frautschi and Suura, who inspected the energy behaviour of infrared (IR) and radiative corrections to generic high-energy processes \cite{AP12379}, and finally extended to gravity by Weinberg in his seminal analysis \cite{PR140B516}. To render transition rates IR finite the scientific community has resorted to sum over all the possible final states compatible with the experiments, a strategy that entails the inclusion of all the possible undetected bosons, i.e. the soft bosons. Nonetheless, even though the transition rate is well defined, the $S$ matrix is still not finite. 

An alternative approach is required to solve the infrared catastrophe with a well defined $S$ matrix. The first attempt in this direction was due to Chung's seminal study \cite{PR140B1110}, where the existence of an alternative representation of the photon states, more appropriate to describe scattering than the usual Fock representation, was shown to exist for QED. Subsequently, Kibble \cite{JMP9315}, Faddeev and Kulish \cite{TMP4745}, and then Ware, Saotome, and R. Akhoury \cite{JHEP2013159} made some important advances on Chung's analysis, developing the dressed state formalism. Within this latter framework, a finite expression for the elements of the $S$ matrix can be recovered by averaging the unitary operator $S$ among dressed states that are introduced as the tensor product of the electronic Fock states and photon coherent states of QED, while satisfying the convergence condition. Furthermore, a framework has been recently developed to deal with infrared divergences, which combines the inclusive probability formalism with the dressed state one \cite{the-third-formalism}.

Since all the observed scattering processes are accompanied by the emission and absorption of soft bosons, one would expect these undetected soft bosons to carry some information. Recently, Carney et al. \cite{PRL119180502,PRD97025007}, and Tomaras et al. \cite{PRD101065006} studied the entanglement entropy of soft photons in the inclusive probability formalism and in the dressed state formalism, by constructing the reduced density matrix for the electron states. Gomez et al. \cite{the-third-formalism} discussed quantum decoherence due to tracing over unobserved radiation in the new combined formalism, and estimated the loss of quantum information due to soft emission. Moreover, Danielson et al. \cite{Danielson-Satishchandran-Wald1,Danielson-Satishchandran-Wald2} showed how the presence of a black hole unavoidably induces soft radiation from the quantum matter on its exterior, and how soft radiation decoheres quantum matter in the same region.

Here, we look at this issue from a different perspective. We aim at quantifying how much information carried by the hard particles in the incoming states can be retrieved by local measurements on the observable outgoing states. This is quantified by the amount of correlations between soft and hard particles. As such, these correlations measure the delocalization of information and are related to the out-of-time-order correlation functions (OTOCs) \cite{kitaev, hosur, shenker}. The delocalization of information, also known as {\em scrambling}, is measured by the tri-partite mutual information $I_3$ \cite{hosur, yoshida}. It is an important feature of quantum chaos \cite{chaos-references-1, chaos-references-2}, black hole physics and the information paradox \cite{fastscrambling, haydenpreskill, kitaevyoshida}. Quantum chaos and its relationship with OTOCs and scrambling in quantum channels and circuits has recently been the subject of interest in the quantum information community \cite{qcic, yoshida,isospectraltwirling, True22}.

In order to compute the scrambling power of soft photons emission, we construct the  Choi state corresponding to the unitary operator $U$ describing the scattering process. We then calculate the tripartite mutual information $I_3$ between the hard particle part of the incoming state, and the hard particle and soft photon part of the outgoing state. This turns out to be a negative quantity  $I_3\le 0$ which vanishes if and only if no scrambling is present. The construction of the matrix expression of the unitary operator $U$ requires both the input and output states to be complete and orthogonal bases. The Fock states satisfy this requirement. Since the $S$ matrix is notoriously not well defined in terms of the Fock states, we turn our focus from the unitary matrix constructed entirely by Fock states to the unitary matrix represented by hard particle Fock states and soft photon coherent states. The unitary operators $S$ that act over these two Hilbert spaces do not coincide, as they differ by a phase operator. Nonetheless, they turn out to be the same operator at the leading order contribution, as for the scattering of a charged particle in an external field \cite{TMP4745}. We can finally proceed to the calculation of the relevant tripartite mutual information in terms of 2-Renyi entropy, and find that this is negative and finite but quite greater than its lower bound - though non maximally scrambling. Since the tripartite mutual information in terms of the 2-Renyi entropy provides an upper bound of the tripartite mutual information in terms of the Von Neumann entropy, we can state that the scattering with soft photon emission-absorption is scrambling and the soft photons carry way relevant information.

The paper is structured as follows. In section \ref{Choi-iso}, we briefly review the Choi isomorphism mapping and how to use Choi states in order to quantify the scrambling of the relevant unitary operator. In section \ref{unitary}, we construct the matrix expression of the unitary operator $U$ in terms of the Fock states, and then focus on the expression in terms of the hard particle Fock states and the soft photon coherent states. Then in section \ref{Choi-matrix} and \ref{tripa}, we provide, respectively, the relevant Choi states and the density matrix, and calculate the relevant tripartite mutual information in terms of 2-Renyi entropy to leading order. In section \ref{dis} we provide a physical interpretation of the result and discuss its profound implication for the black hole information paradox. Finally, in section \ref{conclu} we spell our conclusions and draw an outlook of future analyses.

\section{Choi isomorphism mapping and tripartite mutual information} \label{Choi-iso}
\noindent
In order to develop a quantum-information driven analysis of scattering processes in QFT, specifically in QED, we have to resort to quantum-information measures, which we are able to introduce within the QFT framework by considering a map, dubbed Choi isomorphism map, from a unitary quantum channel to pure states in a doubled Hilbert space \cite{JHEP2016February01}. We denote with $U$ the unitary operator on the quantum channel that can be decomposed in a orthogonal and complete basis $\{\ket{i}\}$ as
\be
U=\sum_{ij}u_{ij}\ket{i}\bra{j}\,.
\ee
Then map this unitary operator $U$ into a doubled state $\ket{U}$, also dubbed Choi state. Its normalized expression is
\be
\label{ref2}
\ket{U} = \frac{1}{\sqrt{D}}\sum_{i,j=0}^{D-1} u_{ji} \ket{i}_{in}\otimes \ket{j}_{out}\,.
\ee
It is also possible to look at this construction in an alternative way. Indeed, we may first define the $D$ dimensional Bell-State $\ket{I}$
\be
\ket{I} = \frac{1}{\sqrt{D}} \sum_{i=0}^{D-1} \ket{i}_{in} \otimes \ket{i}_{out}\,, 
\ee
then the Choi state can be constructed as:
\be 
\label{4}
\ket{U} = (\mathbb{I}_{in}\otimes U_{out})\ket{I}\,.
\ee
$\mathbb{I}_{in}$ is the identity operator acting on the input states. $U_{out}$ is the unitary operator $U$ decomposed on the output states basis, on which it acts. The relevant density matrix can be then expressed as
\be
\label{5}
\rho_U = \ket{U}\bra{U}=(\mathbb{I}_{in}\otimes U_{out})\ket{I} \bra{I}(\mathbb{I}_{in}\otimes U_{out}^{\dag})\,.
\ee
In the what follows we will use this expression of the density matrix to calculate the entropy in some bipartition, and then calculate the tripartite mutual information to quantify the scrambling of the unitary operator $U$.\\

\begin{figure}[h]
\centering
\includegraphics[width=5cm,height=4.5cm]{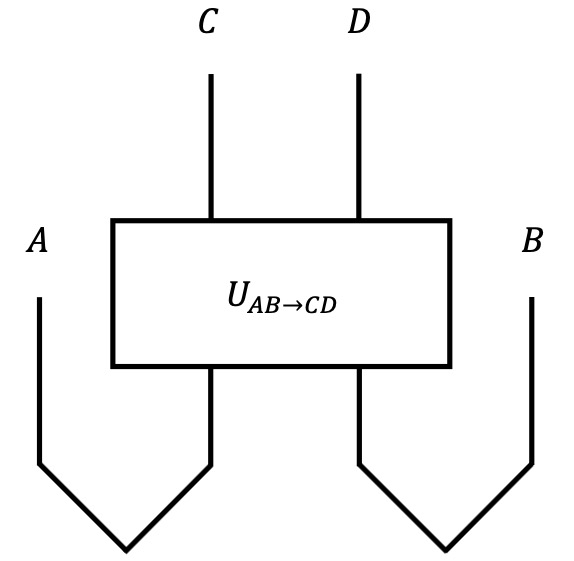}
\caption{Choi state of a bipartite unitary operator $U$.}
\label{figure1}
\end{figure}

The scrambling power is a property of the unitary operator $U$. Since the state defined in $\eqref{ref2}$ contains all the information concerning the inputs and the dynamics of the quantum channel, this can be deployed to establish a general measure for scrambling. At this purpose, we may consider that both the input and output states of the unitary operator $U$ have two degrees of freedom, namely $U=U_{AB \to CD}$, where $A$, $B$ label the two degrees of freedom of the input state, and $C$, $D$ label the two degrees of freedom of the output state, as shown in Fig.~\ref{figure1}. We then map this bipartite unitary operator $U=U_{AB \to CD}$ into a state on the doubled space $\mathcal{H}_A \otimes \mathcal{H}_B \otimes \mathcal{H}_C \otimes \mathcal{H}_D$ --- the $A$, $B$ spaces are the spaces of the input state, the $C$, $D$ spaces are those of the output state --- by means of the Choi isomorphism mapping, which provides the associated Choi state $\ket{U}$. Finally, we are in the position to calculate the tripartite mutual information between the subsystems $A$, $C$, $D$ of the Choi state $\ket{U}$ in terms of the relevant density matrix $\rho_{U}$, which provides a basic measure of the scrambling power on this quantum channel \cite{SPP202110}.\\ 

The definition of the tripartite mutual information is provided by the expression
\be\label{tmi}
I_3(A:C:D)=I(A:C)+I(A:D)-I(A:CD)\,,
\ee
where $I(X:Y)$ is the mutual information between the subsystems $X$, $Y$ of the Choi state $\ket{U}$. This quantity is always less than or equal to zero; the more negative it is, the more scrambling is the process $U$. Thus, the absolute value of the tripartite mutual information accounts for the amount of information about $A$ that are non-locally hidden over $C$ and $D$. It is noteworthy that maximally scrambled information can be recovered from the output exactly because the erasure of correlations between two subsystems in a unitary process needs to be moved somewhere else\cite{haydenpreskill}. Different protocols of information recovery have been proposed, relying either on a perfect knowledge of the unitary process $U$\cite{kitaevyoshida} or, lacking this information, by quantum machine learning\cite{True22, Lloyd23}, provided that the unitary $Y$ is not fully chaotic. It is noteworthy that a unitary process can be fully scrambling but not fully chaotic\cite{chamon23}.

In terms of the von Neumann entropy, the tripartite mutual information casts
\ba
\label{7}
\nn
I_3 &=& S(C) + S(D) - S(AC) - S(AD)\\
&=& \log d_A + \log d_B - S(AC) - S(AD)\,,
\ea
where $S(\rho)$ denotes the von Neumann entropy
\be
S(\rho) = - \tr(\rho\log\rho).
\ee
According to the definition in \eqref{tmi}, the tripartite mutual information is lower- and upper-bounded by \cite{JHEPDecember21}
\be
-2 \log d_A \le I_3 \le 0\,.
\ee
Since the von Neumann entropy is generally hard to be computed, though the analysis we will often focus our attention on the 2-Renyi entropy.\\

The tripartite mutual information in terms of 2-Renyi entropy reads
\ba
\label{10}
\nn
I_{3(2)} &=& S_2(C) + S_2(D) - S_2(AC) - S_2(AD)\\
&=& \log d_A + \log d_B - S_2(AC) - S_2(AD)\,,
\ea
where $S_{2}(\rho)$ denotes the 2-Renyi entropy
\be
S_{2}(\rho) = -\log\tr\rho^2\,.
\ee
Notice that for this latter quantity we also derive the bounds 
\be
-2 \log d_A \le I_{3(2)} \le 0.
\ee

The 2-Renyi entropy, instead of von-Neumann entropy, is appropriate tool in order to estimate the tripartite mutual information that successfully works. This is due to the fact that the tripartite mutual information calculated in terms of the von Neumann entropy is upper-bounded by the tripartite mutual information calculated in terms of the 2-Renyi entropy, namely 
\be
I_3 \le I_{3(2)}\,.
\ee
Thus, if we show that the unitary operator $U$ is scrambling in terms of 2-Renyi entropy, it will be necessarily scrambling also in terms of von Neumann entropy.

\section{Construction of the unitary operator} \label{unitary}
\noindent
The strategy we outlined in the previous section can be deployed to study the scrambling power of the scattering processes with soft photons emission-absorption. In these processes, the input and output states naturally contain two sets of degrees of freedom, hard particles ($A,C$) and soft photons ($B,D$). The evolution operator $S$ that governs this process is unitary, which means there exist a unitary operator $U$ that connects the input states and the output states. Let $\ket{\alpha}_{in}$ and $\ket{\beta}_{out}$ label the input and output states respectively. These are both bases of the Fock space, since each basis of the Fock space is complete and orthogonal, satisfying the condition of constructing the unitary operator $U$. According to quantum mechanics, $\ket{\alpha}_{in}$ and $\ket{\beta}_{out}$ can be connected by a unitary operator $U$ such that \cite{MQM}
\ba\label{bua}
\ket{\beta_i}_{out} = U \ket{\alpha_i}_{in}\,.
\ea
The relation between $U$ and $S$ can be determined according to the relation between $\ket{\alpha}_{in}$ and $\ket{\beta}_{out}$, which reads
\ba\label{asb}
\nn
\ket{\alpha_i}_{in} &=& \sum_j \ket{\beta_j}_{out \ out} \braket{\beta_j | \alpha_i}_{in}=\sum_j S_{ji} \ket{\beta_j}_{out}\\
\nn
&=& \sum_j U \ket{\alpha_j}_{in \ in} \braket{\alpha_j | U^{\dagger} | \alpha_i}_{in}\\
&=& \sum_j \ _{in} \braket{\alpha_j | S | \alpha_i}_{in}  \ket{\beta_j}_{out}\,.
\ea
In our set up, we treat the state ket in the Heisenberg picture as an input state, and deal with the base ket in the Heisenberg picture as an output state. Comparing \eqref{bua} with \eqref{asb}, one can obtain that the relation between $U$ and $S$, i.e.
\be
U = S^{\dagger}\,.
\ee
It is straightforward to construct the unitary operator $U$ in terms of the input and output states, and decompose it in terms of the output states, namely 
\ba
\nn
U &=& \sum_l \ket{\beta_l}_{out \ in} \bra{\alpha_l} = \sum_{l} U \ket{\alpha_l}_{in \ in} \bra{\alpha_l}\\
\nn
&=& \sum_{ijl} \ket{\beta_i}_{out \ out} \braket{\beta_i | \beta_l}_{out \ in} \braket{\alpha_l | \beta_j}_{out \ out} \bra{\beta_j}\\
&=& \sum_{ij} S^*_{ji} \ket{\beta_i}_{out \ out} \bra{\beta_j}\,.
\ea
Then the relevant Choi state can be expressed as
\ba
\nn
\ket{U} &=& (\mathbb{I}_{in}\otimes U_{out})\ket{I} = \frac{1}{\sqrt{D}}\sum_{i=0}^{D-1} \ket{\alpha_i}_{in}\otimes U \ket{\beta_i}_{out}\\
&=& \frac{1}{\sqrt{D}}\sum_{i,j=0}^{D-1} S^*_{ij} \ket{\alpha_i}_{in}\otimes \ket{\beta_j}_{out}\,.
\ea

\begin{figure}[h]
\centering
\includegraphics[width=8.25cm,height=3.25cm]{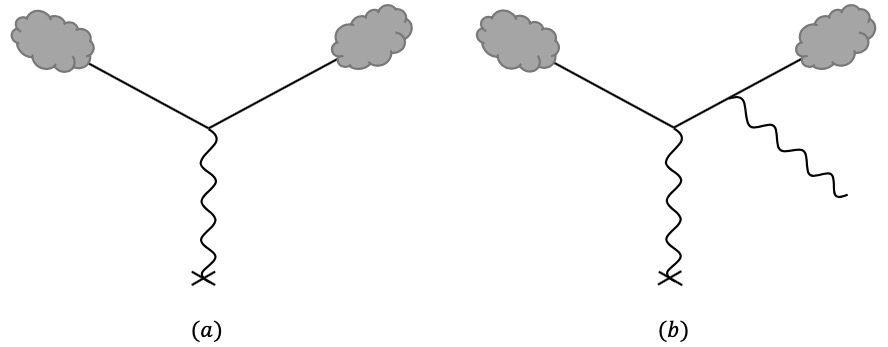}
\caption{Type (a) contributions are leading-order, while type (b) are sub-leading-order contributions.}
\label{figure2}
\end{figure}

In perturbative QED, the unitary operator $S$ can be expanded around the identity $\mathbb{I}$, with infinitely many terms of the expansion. At an operative level, it is unrealistic to retain all the terms of expansion of $S$, as the sensitivity of the measurements will automatically pick out the order of the perturbative expansion at which the theory can be confronted with the experiments. For this reason, we will restrict our focus on leading order terms in the expansion. The evolution operator is still unitary in this approximation, so the aforementioned method is still valid. We may consider the paradigmatic study case of an electron scattered by an external electromagnetic field. Since all observed scattering processes in QED are accompanied by emission-absorption of photons \cite{PR5254,AP12379,QFTMS}, this case represents a valuable example to study the scrambling power of electrons and soft photons. In this case, the incoming state (or the input state) is represented by
\be
\label{ref19}
\ket{\alpha}_{in} = \ket{p_i, \{k_j^{(\lambda_j)}\}_{n_j}}_{in}\,,
\ee
where $p$ denotes the momentum of the electron, $k$, $\lambda$ and $n$ label respectively the momentum, polarization and occupation number of photon. The subscripts $i$, $j$ are used to distinguish the different quantities involved, namely $p$, $k$, $\lambda$ and $n$. Therefore, the expression 
\be
\ket{\{k_j^{(\lambda_j)}\}_{n_j}}_{in} = \ket{\{n_j(k_j^{(\lambda_j)})\}}_{in} = \prod_j \ket{n_j(k_j^{(\lambda_j)})}_{in}
\ee
indicates that there are $\{n_j\}$ photons with momentum $\{k_j\}$ and polarization $\{\lambda_j\}$. Similarly, the outgoing state (or the output state) is
\be
\label{ref21}
\ket{\beta}_{out} = \ket{p_i, \{k_j^{(\lambda_j)}\}_{n_j}}_{out} = U \ket{p_i, \{k_j^{(\lambda_j)}\}_{n_j}}_{in}\,.
\ee
Then the unitary operator can be constructed as
\ba
\label{ref22}
\nn
U &=& \sum_{\{p_i\}} \sum_{\{n_j\}} \ket{p_i, \{k_j^{(\lambda_j)}\}_{n_j}}_{out \ in} \bra{p_i, \{k_j^{(\lambda_j)}\}_{n_j}}\\
\nn
&=& \sum_{\{p_i\}} \sum_{\{n_j\}} U \ket{p_i, \{k_j^{(\lambda_j)}\}_{n_j}}_{in \ in} \bra{p_i, \{k_j^{(\lambda_j)}\}_{n_j}}\\
&=& \sum_{\{n_j\}} \int d^3p_i \ U \ket{p_i, \{k_j^{(\lambda_j)}\}_{n_j}}_{in \ in} \bra{p_i, \{k_j^{(\lambda_j)}\}_{n_j}}\,. \qquad
\ea
In the latter two passages of \eqref{ref22}, we switched for convenience from a regularization of the state momenta in a finite fiducial box, to a representation of the momenta in the continuum limit --- see also Appendix \ref{A-C}. The relevant completeness relations are
\ba
&& \int d^3p \ \ket{p}_{in \ in} \bra{p} = \mathbb{I}_A.\\
&& \sum_{\{n_j\}} \ket{\{k_j^{(\lambda_j)}\}_{n_j}}_{in \ in} \bra{ \{k_j^{(\lambda_j)}\}_{n_j}} = \mathbb{I}_B.\\
&& \sum_{\{n_j\}} \int d^3p_i \ \ket{p_i, \{k_j^{(\lambda_j)}\}_{n_j}}_{in \ in} \bra{p_i, \{k_j^{(\lambda_j)}\}_{n_j}} = \mathbb{I}\,. \quad
\ea
Scattering processes are in general accompanied by the emission-absorption of hard photons, nonetheless this can be regarded as a sub-leading order contribution, and therefore ignored. Since Fock states encode information of the occupation number of particle states with different momenta, it is operationally meaningful to separate hard particles (electrons and hard photons) from the soft particles (soft photons) as two different sets of degrees of freedom. According to the dressed state formalism, soft photons accompanying charged particles can be regarded as clouds surrounding the charged particles, thus as an intrinsic property of these latter ones. Being scattering processes involving hard photons also accompanied by soft photons, scattering process involving hard photons, once compared with the scattering processes without hard photons, can be considered as sub-leading order contribution, therefore ignored. The relevant schematic diagram is depicted in Figure \ref{figure2}. Both the leading order contribution (a) and the sub-leading order contribution (b) can be in principle taken into account. Nonetheless, at the first order in the perturbative expansion of the $S$ matrix, only diagrams of the type (a) contribute, which enables to consider only soft photons processes.

While in our analysis we move essentially from a complete basis of a subspace of the Fock space, a viable alternative is to start from the complete basis of the whole Fock space, and then retain only the leading-order of the final results. It is worth noting that, within the second case possibility, the unitary operator $S$ within the dressed state formalism is not exactly the same as the unitary operator within the inclusive probability formalism, as this latter one differs by a phase operator \cite{TMP4745}, and that some additional operations would be then required. Nonetheless, when we consider the leading order contribution, charged particles scattered by an external field, the phase operator vanishes. Thus, at least for this exemplified but still paradigmatic case, the two methods deployed in order to calculate the components of the unitary operator $S$ will provide the same results, and thus the same leading-order expansion.\\

We have hitherto construct the unitary operator matrix elements, within the Fock states basis. Nonetheless, since the $S$ matrix elements that contain soft photons both in emission and absorption are not well defined, we shall resort to a transform from the states of the photon Fock space to the coherent states, which finally allow to define the $S$ matrix elements. Accordingly, for the space on which the unitary operator $S$ acts we consider 
\cite{TMP4745}
\be
\mathcal{H}_{FQ} \otimes \mathcal{H}_{F\gamma} \to \mathcal{H}_{FQ} \otimes \mathcal{H}_{C\gamma}\,,
\ee
where $\mathcal{H}_{FQ}$ is the electron Fock states space, $\mathcal{H}_{F\gamma}$ is the photon Fock states space, and $\mathcal{H}_{C\gamma}$ is the photon coherent states space.\\ 

We can not construct the Choi state $\ket{U}$ directly from the states that belong to $\mathcal{H}_{FQ} \otimes \mathcal{H}_{C\gamma}$, because the states within this space are not orthogonal. Nonetheless, we can always use the completeness relation to obtain the expression of the Choi state in terms of the states that are in $\mathcal{H}_{FQ} \otimes \mathcal{H}_{C\gamma}$.\\

The definition of the photon coherent state reads \cite{PR140B1110}
\ba \label{csp}
\nn
\ket{\beta_{a}^l} &=& D(\beta_{a}^l) \ket{0} = \exp \Big[\beta_{a}^l a_{a}^{l{\dagger}} - \beta_{a}^{l*} a_{a}^{l}\Big] \ket{0}\\
\nn
&=& \text{exp}\Big(-\frac{1}{2} |\beta_{a}^l|^2\Big) \text{exp}\Big(\beta_{a}^l a_{a}^{l{\dagger}}\Big) \ket{0}\\
&=& \text{exp}\Big(-\frac{1}{2} |\beta_{a}^l|^2\Big) \sum_{n} \frac{(\beta_{a}^l)^n}{\sqrt{n!}} \ket{n_{l,a}}\,,
\ea
with
\be \label{daggerf}
a_{a}^{l{\dagger}} = \int d^3k f_a(k) a^{(l){\dagger}}(k)\,.
\ee
In \eqref{daggerf} the set of functions $f_a(k)$ is complete and orthonormal, and is defined on some region of the momentum space that includes $k=0$. The ladder operator $a^{(l){\dagger}}(k)$ in \eqref{csp} denotes the creation operator of a state of the photon Fock space; acting on the vacuum, it creates a photon with momentum $k$ and polarization $l$. The ladder operator $a_{a}^{l{\dagger}}$ is also a creation operator for \eqref{csp} that, acting on the vacuum, creates a photon with momentum distribution $f_{a}(k)$ and polarization $l$. Finally, $\ket{n_{l,a}}$ represents a generic number state for $n_{l,a}$ photons with momentum distribution $f_{a}(k)$ and polarization $l$. The tensor product over all the possible $l$ and $a$ of a generic state $\ket{\beta_{a}^l}$ provides the generic photon coherent state
\ba
\nn
\ket{\{\beta_{a}^l\}} &=& \prod_{l,a} \ket{\beta_{a}^l} = D(\{\beta_{a}^l\}) \ket{0}\\
&=& \exp \Big[\sum_{l,a} \beta_{a}^l a_{a}^{l{\dagger}} - \beta_{a}^{l*} a_{a}^{l}\Big] \ket{0}\,.
\ea
Thus, the states that belong to $\mathcal{H}_{FQ} \otimes \mathcal{H}_{C\gamma}$ are expressed by 
\be
\ket{p_i,\{\beta_{ia}^l\}} \in \mathcal{H}_{FQ} \otimes \mathcal{H}_{C\gamma}\,,
\ee
where the subscript $i$ is used to distinguish different $p$ and $\{\beta_{a}^l\}$. Assuming the total number of $l,a$ to be $N$, completeness relations cast
\ba
&& \Big(\frac{1}{\pi}\Big)^N \int d^2\{\beta_{ia}^{l}\} \ \ket{\{\beta_{ia}^{l}\}} \bra{\{\beta_{ia}^{l}\}} = \mathbb{I}_{B}\,,\\
\label{ref32}
&& \Big(\frac{1}{\pi}\Big)^N \int d^3p_i \int d^2\{\beta_{ia}^{l}\} \ \ket{p_i, \{\beta_{ia}^{l}\}} \bra{p_i, \{\beta_{ia}^{l}\}} = \mathbb{I}\,.\quad
\ea
Similarly as in \eqref{ref19} and \eqref{ref21}, we can look at the state $\ket{p_i,\{\beta_{ia}^l\}}$ as an incoming state, and then consider the state that we obtain by applying the unitary operator $U$ on the state $\ket{p_i,\{\beta_{ia}^l\}}$ as the outgoing state, namely 
\be
\ket{\alpha}_{in} = \ket{p_i, \{\beta_{ia}^{l}\}}, \ \ket{\beta}_{out} = U \ket{p_i, \{\beta_{ia}^{l}\}}\,.
\ee
With this definition, we recover the completeness relation for the outgoing states by $U \mathbb{I} U^{\dagger} = \mathbb{I}$. Inserting it into \eqref{ref22}, an expression for $U$ in terms of the outgoing states can be then achieved, which is equivalent to applying $U$ to the identity in \eqref{ref32}, decomposing the result in terms of the outgoing states. What we finally recover is nothing but $U_{out}$, which is essential to construct the Choi state $\ket{U}$.\\

\begin{widetext}
\ba
\label{35}
\nn
U &=& \sum_{\{m_j\}} \int d^3p_i \Big[ \Big(\frac{1}{\pi}\Big)^N \int d^3p_k \int d^2\{\beta_{ka}^{l}\} \ U \ket{p_k, \{\beta_{ka}^{l}\}} \bra{p_k, \{\beta_{ka}^{l}\}} U^{\dagger} \Big] \Big[ U \ket{p_i, \{k_j^{(\lambda_j)}\}_{m_j}}_{in \ in} \bra{p_i, \{k_j^{(\lambda_j)}\}_{m_j}} \Big]\\
\nn
&& \Big[ \Big(\frac{1}{\pi}\Big)^N \int d^3p_l \int d^2\{\beta_{la}^{l}\} \ U \ket{p_l, \{\beta_{la}^{l}\}} \bra{p_l, \{\beta_{la}^{l}\}} U^{\dagger} \Big]\\
\nn
&=& \Big(\frac{1}{\pi}\Big)^{3N} \int d^3p_i d^3p_l \int d^2\{\beta_{ka}^{l}\} d^2\{\beta_{la}^{l}\} d^2\{\beta_{ha}^{l}\} \braket{\{\beta_{ka}^{l}\} | \{\beta_{ha}^{l}\}}  \bra{p_i, \{\beta_{ha}^{l}\}} S^{\dagger} \ket{p_l, \{\beta_{la}^{l}\}} \ U \ket{p_i, \{\beta_{ka}^{l}\}}  \bra{p_l, \{\beta_{la}^{l}\}} U^{\dagger}\,,\\
\ea
where we made use of the relation 
\ba
&& \bra{p_k, \{\beta_{ka}^{l}\}} U^{\dagger}U \ket{p_i, \{k_j^{(\lambda_j)}\}_{m_j}}_{in} = \braket{p_k, \{\beta_{ka}^{l}\} | p_i, \{k_j^{(\lambda_j)}\}_{m_j}}_{in} = \delta^{(3)} (p_i-p_k) \braket{\{\beta_{ka}^{l}\} | \{k_j^{(\lambda_j)}\}_{m_j}}_{in}\,,
\ea
and
\ba
\nn
&& _{in} \bra{p_i, \{k_j^{(\lambda_j)}\}_{m_j}} U \ket{p_l, \{\beta_{la}^{l}\}} = \Big(\frac{1}{\pi}\Big)^N \int d^3p_h \int d^2\{\beta_{ha}^{l}\}  \ _{in} \braket{p_i, \{k_j^{(\lambda_j)}\}_{m_j} | p_h, \{\beta_{ha}^{l}\}} \bra{p_h, \{\beta_{ha}^{l}\}} U \ket{p_l, \{\beta_{la}^{l}\}}\\
&& = \Big(\frac{1}{\pi}\Big)^N \int d^3p_h \int d^2\{\beta_{ha}^{l}\}  \ _{in} \delta^{(3)}(p_i-p_h) \braket{\{k_j^{(\lambda_j)}\}_{m_j} | \{\beta_{ha}^{l}\}} \bra{p_h, \{\beta_{ha}^{l}\}} S^{\dagger} \ket{p_l, \{\beta_{la}^{l}\}}\,.
\ea
\end{widetext}

In principle, $\ket{p_i,\{\beta_{ia}^l\}}$ does not necessarily represent the dressed state corresponding to $\ket{p_i}$. To construct the unitary operator $U$ we rather need the complete basis of the space $\mathcal{H}_{FQ} \otimes \mathcal{H}_{C\gamma}$. Nonetheless, not every state $\ket{p_i,\{\beta_{ia}^l\}}$ within this space is a dressed state corresponding to $\ket{p_i}$. We still need to figure out these states have to be dealt with in our calculations. At this purpose, we consider that dressed states can be extracted by inspecting the expression of the unitary operator $\exp\{R_f\}$, which was provided by Kulish and Faddeev in Ref.~\cite{TMP4745}, i.e.
\ba
\nn
&& \exp\{R_f\} = \exp \bigg[ \frac{e}{(2\pi)^{3/2}} \int \sum_{l=1}^{2} \big( F^{(l)}(k,p) a^{(l){\dagger}}(k)\\
&& \qquad \qquad \quad \ -  F^{(l)*}(k,p) a^{(l)}(k) \big) \rho(p) d^3p \frac{d^3k}{(2k^0)^{1/2}} \bigg]\,, \qquad
\ea
where
\ba
&& F^{(l)}(k,p) = \frac{p \cdot \epsilon^{(l)}(k)}{p \cdot k} \phi(k,p)\,,\\
&& \rho(p) = \sum_{n} b_n^{\dagger}(p)b_n(p) - d_n^{\dagger}(p)d_n(p)\,,
\ea
with $\phi(k,p)$ a smooth function that satisfies $\phi(k,p) \to 1$ as $k \to 0$. The dressed state created by $\exp\{R_f\}$ are then recovered to be
\ba
\exp\{R_f\} \ket{p_i} &=& \ket{p_i, \{\beta_{a}^{l}(p_i)\}}\\
\nn
&=& \ket{p_i} \otimes D(\beta_{a}^{l}(p_i)) \ket{0}\,, \quad
\ea
with the displacement operator $D(\beta_{a}^{l}(p_i))$ casts in turn as
\be
D(\beta_{a}^{l}(p_i)) = \exp \Big[\sum_{l,a} \beta_{a}^l(p_i) a_{a}^{l{\dagger}} - \beta_{a}^{l*}(p_i) a_{a}^{l}\Big]\,.
\ee
The state $\ket{\{\beta_{a}^{l}(p_i)\}}$ generated by the action of $D(\beta_{a}^{l}(p_i))$ is expressed as
\ba
\nn
&& \ket{\{\beta_{a}^l(p_i)\}} = \exp \Big[\sum_{l,a} \beta_{a}^l(p_i) a_{a}^{l{\dagger}} - \beta_{a}^{l*}(p_i) a_{a}^{l}\Big] \ket{0}\\
\nn
&& = \text{exp}\Big\{-\frac{1}{2} \sum_l \int d^3k |\tilde{S}_i^{(l)}(k)|^2\Big\}\\
\nn
&& \qquad \times \text{exp} \Big\{\sum_l \int d^3k \tilde{S}_i^{(l)}(k) a^{(l){\dagger}}(k)\Big\} \ket{0}\\
\nn
&& = \text{exp}\Big(-\frac{1}{2}\sum_{l,a} |\beta_{a}^l(p_i)|^2\Big)\\
&& \qquad \times \text{exp}\Big(\sum_{l,a}\beta_{a}^l(p_i) \int d^3k f_a(k) a^{(l){\dagger}}(k)\Big) \ket{0}\,,
\ea
where $\tilde{S}_i^{(l)}(k)$ denotes the usual soft factor --- since the photon coherent state corresponds here to the soft photon coherent state, it is safe to write $\phi(k,p)=1$ --- that reads
\ba
\tilde{S}_i^{(l)}(k) = \frac{e}{[2(2\pi)^3k_0]^{1/2}} \frac{p_i \cdot \epsilon^{(l)}(k)}{k \cdot p_i} = \sum_a \beta_{a}^l(p_i) f_a(k)\,. \qquad
\ea
Accordingly, the state $\ket{p_i, \{\beta_{ia}^{l}\}}$ in $\mathcal{H}_{FQ} \otimes \mathcal{H}_{C\gamma}$ individuates a dressed state corresponding to $\ket{p_i}$ if the momentum-dependence is introduced in $\{\beta_{ia}^{l}\} = \{\beta_{a}^{l}(p_i)\}$, which is a necessary condition to for the convergence condition $\sum_{l,a} |\beta^{l}_{a}|^2 < \infty$ to be realized \cite{PR140B1110,TMP4745,PRD104125004}. We may wonder how matrix elements of the form $\braket{p_{i'}, \{\beta_{i'a}^{l}\} | S | p_{i}, \{\beta_{ia}^{l}\}}$ can be managed. Chung mentioned that complex coefficients $\{\beta_{ia}^{l}\}$ and $\{\beta_{i'a}^{l}\}$, which specify the initial and final coherent states of soft photon, could be written as \cite{PR140B1110}
\ba \label{betas}
\beta_{ia}^{l} = \beta_{a}^{l}(p_i) + \epsilon_{ia}^{l}, \qquad \beta_{i'a}^{l} = \beta_{a}^{l}(p_{i'}) + \epsilon_{i'a}^{l}\,.
\ea
Using \eqref{betas}, it is possible to show that the matrix elements $\braket{p_{i'}, \{\beta_{i'a}^{l}\} | S | p_{i}, \{\beta_{ia}^{l}\}}$ are IR finite and computable, i.e.
\ba
\label{46}
\nn
&& \braket{p_{i'}, \{\beta_{i'a}^{l}\} | S | p_{i}, \{\beta_{ia}^{l}\}} = \braket{p_{i'}, \{\beta_{i'a}^{l}\} | \mathbb{I} - i T | p_{i}, \{\beta_{ia}^{l}\}}\\[1mm]
\nn
&& = \delta^{(3)}(p_{i'} - p_i) \braket{\{\beta_{i'a}^{l}\} | \{\beta_{ia}^{l}\}}\\[1mm]
\nn
&& \qquad \quad - 2i\pi \delta^{(4)}(p_{i'} - p_i) \braket{p_{i'}, \{\beta_{i'a}^{l}\} | M | p_{i}, \{\beta_{ia}^{l}\}}\\[1mm]
\nn
&& \simeq \delta^{(3)}(p_{i'} - p_i) \braket{\{\beta_{i'a}^{l}\} | \{\beta_{ia}^{l}\}}\\[1mm]
&& \qquad \quad - 2i\pi \delta(p_{i'}^0 - p_i^0) \braket{p_{i'}, \{\beta_{i'a}^{l}\} | M_e | p_{i}, \{\beta_{ia}^{l}\}}\,,
\ea
where $\braket{p_{i'}, \{\beta_{i'a}^{l}\} | M | p_{i}, \{\beta_{ia}^{l}\}}$ is a delta-function-free matrix, also called $M$ matrix --- in other words, the $M$ matrix is recovered by removing delta-function contributions from the matrix $\braket{p_{i'}, \{\beta_{i'a}^{l}\} | T | p_{i}, \{\beta_{ia}^{l}\}}$. The quantity  $\braket{p_{i'}, \{\beta_{i'a}^{l}\} | M_e | p_{i}, \{\beta_{ia}^{l}\}}$ denotes the leading order contribution of $\braket{p_{i'}, \{\beta_{i'a}^{l}\} | M | p_{i}, \{\beta_{ia}^{l}\}}$. This is nothing but the $M$ matrix for the process involving a charged particle scattered by an external field \cite{PR140B1110}, and reads
\ba
\label{47}
\nn
&& \braket{p_{i'}, \{\beta_{i'a}^{l}\} | M_e | p_{i}, \{\beta_{ia}^{l}\}}\\
\nn
&& = \exp (\alpha B (p_i, p_{i'}) + \alpha \tilde{B} (p_i, p_{i'})) \exp \Big\{ \sum_{l,a} \big[ - \frac{1}{2}|\epsilon_{i'a}^{l} -  \epsilon_{ia}^{l}|^2\\
\nn
&& + i \text{Im} ( \beta_{a}^{l*}(p_i) \epsilon_{ia}^{l} + \beta_{a}^{l}(p_{i'}) \epsilon_{i'a}^{l*} - \beta_{a}^{l}(p_i)  \beta_{a}^{l*}(p_{i'}) + \epsilon_{ia}^{l} \epsilon_{i'a}^{l*}) \big] \Big\}\\
&&  \Big[\sum_{m,m'=0}^{\infty} m_{m,m'} (p_{i}, p_{i'}) \Big]\,.
\ea
The two functions $B (p_i, p_{i'})$ and $\tilde{B} (p_i, p_{i'})$ share the same order of divergence, but have opposite sign, thus the argument of the first exponential is free from infrared divergences when the regulator is removed, i.e. in the zero limit mass for the photon. The other terms are as well free from infrared divergences.\\

Whenever the states satisfy the condition
\ba
\sum_{l,a} |\epsilon_{i'a}^{l} -  \epsilon_{ia}^{l}|^2 \sim \infty,
\ea
contributions to the processes' amplitude that arise from  $\braket{p_{i'}, \{\beta_{i'a}^{l}\} | M_e | p_{i}, \{\beta_{ia}^{l}\}}$ are vanishing. Notice however that even if the condition 
\ba
\sum_{l,a} |\epsilon_{i'a}^{l} -  \epsilon_{ia}^{l}|^2 < \infty
\ea
is satisfied, then $\braket{p_{i'}, \{\beta_{i'a}^{l}\} | M_e | p_{i}, \{\beta_{ia}^{l}\}}$ is still free of infrared divergence and non-vanishing.

\begin{widetext}
\section{Density matrix and tripartite mutual information expressed in terms of the Choi states} \label{Choi-matrix}
\noindent
In the previous section we expressed the unitary operator $U$ in terms of the states in $\mathcal{H}_{FQ} \otimes \mathcal{H}_{C\gamma}$, more precisely in terms of the outgoing state in $\mathcal{H}_{FQ} \otimes \mathcal{H}_{C\gamma}$). 
We now set out to compute the reduced density operators necessary to the computation of $I_3$. Calculations are heavy and cumbersome and details can be found in the Appendix.
Starting with the expression of the Choi states we have 
\ba \label{Choi-ce}
\nn
&& \ket{U} = (I_{in} \otimes U_{out}) \ket{I} = \frac{1}{\sqrt{D}} \sum_{\{n_{j'}\}} \int d^3p_{i'} \Big( \ket{p_{i'}, \{k_{j'}^{(\lambda_j')}\}_{n_{j'}}} \Big) \otimes U \Big( U \ket{p_{i'}, \{k_{j'}^{(\lambda_j')}\}_{n_{j'}}} \Big)\\
\nn
&& = \frac{1}{\sqrt{D}} \Big(\frac{1}{\pi}\Big)^{4N} \sum_{\{n_{j'}\}} \int d^3p_{i'} d^3p_i \int d^2\{\beta_{ga}^{l}\} d^2\{\beta_{ka}^{l}\} d^2\{\beta_{la}^{l}\} d^2\{\beta_{ha}^{l}\} \braket{\{\beta_{ga}^{l}\} | \{k_{j'}^{(\lambda_j')}\}_{n_{j'}}} \braket{\{\beta_{ka}^{l}\} | \{\beta_{ha}^{l}\}} \braket{\{\beta_{la}^{l}\} | \{k_{j'}^{(\lambda_j')}\}_{n_{j'}}}\\
&& \bra{p_i, \{\beta_{ha}^{l}\}} S^{\dagger} \ket{p_{i'}, \{\beta_{la}^{l}\}} \ \ket{p_{i'}, \{\beta_{ga}^{l}\}} \otimes U \ket{p_i, \{\beta_{ka}^{l}\}}\,.
\ea
The associated density matrix  can be thus written as
\ba
\nn
&& \rho_U = \ket{U} \bra{U}\\
\nn
&& = \frac{1}{D} \Big(\frac{1}{\pi}\Big)^{8N} \sum_{\{n_{j'}\}} \sum_{\{m_j\}} \int d^3p_{i'} d^3p_i d^3p_{r'} d^3p_r \int d^2\{\beta_{ga}^{l}\} d^2\{\beta_{ka}^{l}\} d^2\{\beta_{la}^{l}\} d^2\{\beta_{ha}^{l}\} d^2\{\beta_{g'a}^{l}\} d^2\{\beta_{k'a}^{l}\} d^2\{\beta_{l'a}^{l}\} d^2\{\beta_{h'a}^{l}\} \quad \ \\
\nn
&& \braket{\{\beta_{ga}^{l}\} | \{k_{j'}^{(\lambda_j')}\}_{n_{j'}}}_{in} \braket{\{\beta_{ka}^{l}\} | \{\beta_{ha}^{l}\}}  \bra{p_i, \{\beta_{ha}^{l}\}} S^{\dagger} \ket{p_{i'}, \{\beta_{la}^{l}\}} \braket{\{\beta_{la}^{l}\} | \{k_{j'}^{(\lambda_j')}\}_{n_{j'}}}_{in \ in} \braket{ \{k_{j}^{(\lambda_j)}\}_{m_j} | \{\beta_{g'a}^{l}\}} \braket{\{\beta_{h'a}^{l}\} | \{\beta_{k'a}^{l}\}}\\
&& \bra{p_{r'}, \{\beta_{l'a}^{l}\}} S \ket{p_r, \{\beta_{h'a}^{l}\}} \ _{in} \braket{\{k_{j}^{(\lambda_j)}\}_{m_j} | \{\beta_{l'a}^{l}\}} \ket{p_{i'}, \{\beta_{ga}^{l}\}} \bra{p_{r'}, \{\beta_{g'a}^{l}\}} \otimes \ U \ket{p_i, \{\beta_{ka}^{l}\}} \bra{p_r, \{\beta_{k'a}^{l}\}} U^{\dagger}\,.
\ea
In order to calculate the tripartite mutual information $I_{3(2)}$ in equation \eqref{10}, we have to construct the relevant reduced density matrix related to the states in \eqref{Choi-ce}.
The density matrix $\rho_U$ is essentially represented in terms of the states in $\mathcal{H}_{FQ} \otimes \mathcal{H}_{F\gamma}$. It can be further recast in terms of the states in $\mathcal{H}_{FQ} \otimes \mathcal{H}_{C\gamma}$. Finally, tracing out the relevant states in $\mathcal{H}_{FQ} \otimes \mathcal{H}_{F\gamma}$, rather than the states in $\mathcal{H}_{FQ} \otimes \mathcal{H}_{C\gamma}$, the following reduced density matrices, essential to our strategy, can be attained. Denoting with $A$ and $C$, respectively, the hard components of the incoming states, and outgoing states. And similarly denoting with $B$ and $D$, respectively, the soft components of the incoming states, and outgoing states. Then the states in $\mathcal{H}_{FQ} \otimes \mathcal{H}_{F\gamma}$, more precisely, the states in \eqref{ref19} and \eqref{ref21} can be labeled
\ba
&& \ket{\alpha}_{in} = \ket{p_i, \{k_j^{(\lambda_j)}\}_{n_j}}_{in} = \ket{p_A,\{k_B\}},\\
&& \ket{\beta}_{out} = \ket{p_i, \{k_j^{(\lambda_j)}\}_{n_j}}_{out} = \ket{p_C,\{k_D\}}.
\ea

The reduced density matrix of $A$ and $C$ states reads, after a calculation,
\ba
\nn
&& \rho_{AC} = \tr_{BD} \rho_U\\
\nn
&& = \frac{1}{D} \Big(\frac{1}{\pi}\Big)^{2N} \int d^3p_{i'} d^3p_i d^3p_{r'} d^3p_{r} \int d^2\{\beta_{i'a}^{l}\} d^2\{\beta_{ia}^{l}\} \braket{p_i, \{\beta_{ia}^{l}\} | S^{\dagger} | p_{i'}, \{\beta_{i'a}^{l}\}} \braket{p_{r'}, \{\beta_{i'a}^{l}\} | S | p_r, \{\beta_{ia}^{l}\}} \ \ket{p_{i'}}_{in \ in} \bra{p_{r'}} \quad\\
&& \otimes \ \ket{p_{i}}_{out \ out} \bra{p_{r}}\,,
\ea
where states $B$, which denote the soft component of the incoming states, and states $D$, the outgoing states, have been traced out.

The reduced density matrix of the $C$ states is  recovered by tracing out the hard part components $A$ of the incoming states, 
\ba
\nn
&& \rho_{C} = \tr_{A} \rho_{AC}\\
&& = \frac{1}{D} \Big(\frac{1}{\pi}\Big)^{N} \int d^3p_i d^3p_{r} \int d^2\{\beta_{ia}^{l}\} \braket{\{\beta_{ia}^{l}\} | \{\beta_{ia}^{l}\}} \delta^{(3)}(p_{i} - p_{r}) \ \ket{p_{i}}_{out \ out} \bra{p_{r}} = \frac{d_{B}}{D} \int d^3p_i \ \ket{p_{i}}_{out \ out} \bra{p_{i}}. \qquad \qquad \qquad
\ea

Similarly, tracing out $B$ and $C$ states, respectively ongoing soft photons and incoming hard part charged particle, we obtain for the density matrix related to incoming hard charged particles $A$ and outgoing soft photons $D$
\ba
\nn
&& \rho_{AD} = \tr_{BC} \rho_U\\
\nn
&& = \frac{1}{D} \Big(\frac{1}{\pi}\Big)^{3N} \sum_{\{m'_{s'}\}} \sum_{\{m''_{s''}\}} \int d^3p_{i'} d^3p_i d^3p_{r'} \int d^2\{\beta_{i'a}^{l}\} d^2\{\beta_{ia}^{l}\} d^2\{\beta_{h'a}^{l}\} \braket{p_i, \{\beta_{ia}^{l}\} | S^{\dagger} | p_{i'}, \{\beta_{i'a}^{l}\}} \braket{p_{r'}, \{\beta_{i'a}^{l}\} | S | p_i, \{\beta_{h'a}^{l}\}}\\
&& \braket{\{\beta_{h'a}^{l}\} | \{k_{s''}^{(\lambda_{s''})}\}_{m''}}_{in \ in} \braket{\{k_{s'}^{(\lambda_{s'})}\}_{m'} | \{\beta_{ia}^{l}\}} \ \ket{p_{i'}}_{in \ in} \bra{p_{r'}} \otimes \ket{\{k_{s'}^{(\lambda_{s'})}\}_{m'}}_{out \ out} \bra{\{k_{s''}^{(\lambda_{s''})}\}_{m''}}.
\ea
Now, tracing out incoming charged particle $A$, we derive for outgoing soft particles $D$
\ba
 \rho_{D} = \tr_{A} \rho_{AD}
 = \frac{d_{A}}{D} \sum_{\{m'_{s'}\}} \ket{\{k_{s'}^{(\lambda_{s'})}\}_{m'_{s'}}}_{out \ out} \bra{\{k_{s'}^{(\lambda_{s'})}\}_{m'_{s'}}}. \qquad \qquad \qquad \qquad \qquad \qquad \qquad \qquad \qquad \qquad \qquad \qquad \quad
\ea
It is now straightforward to compute the 2-Renyi entropies related to the aforementioned quantities:
\ba
\nn
S_2(AC) &=& -\log\tr\rho_{AC}^2\\
\nn
&=& -\log \Big[ \frac{1}{D^2} \Big(\frac{1}{\pi}\Big)^{8N} \int d^3p_{i'} d^3p_i d^3p_{r'} d^3p_r \int d^2\{\beta_{i'a}^{l}\} d^2\{\beta_{ia}^{l}\} d^2\{\beta_{r'a}^{l}\} d^2\{\beta_{ra}^{l}\} d^2\{\beta_{g'a}^{l}\} d^2\{\beta_{ga}^{l}\} d^2\{\beta_{h'a}^{l}\} d^2\{\beta_{ha}^{l}\} \\
\nn
&& \braket{p_i, \{\beta_{ia}^{l}\} | S^{\dagger} | p_{i'}, \{\beta_{i'a}^{l}\}} \braket{\{\beta_{i'a}^{l}\} | \{\beta_{r'a}^{l}\}} \braket{p_{r'}, \{\beta_{r'a}^{l}\} | S | p_r, \{\beta_{ra}^{l}\}} \braket{\{\beta_{ra}^{l}\} | \{\beta_{ia}^{l}\}} \braket{p_r, \{\beta_{ga}^{l}\} | S^{\dagger} | p_{r'}, \{\beta_{g'a}^{l}\}}\\
&& \braket{\{\beta_{g'a}^{l}\} | \{\beta_{h'a}^{l}\}} \braket{p_{i'}, \{\beta_{h'a}^{l}\} | S | p_i, \{\beta_{ha}^{l}\}} \braket{\{\beta_{ha}^{l}\} | \{\beta_{ga}^{l}\}} \Big];
\ea
\be
S_2(C) = -\log\tr\rho_{C}^2 = -\log\tr \Big[ \frac{d_{B}^2}{D^2} \int d^3p_i \ \ket{p_{i}}_{out \ out} \bra{p_{i}} \Big] = -\log \Big[ \frac{d_{B}^2d_{A}}{D^2} \Big] = \log d_{A}\,; \qquad \qquad \qquad \qquad \quad \quad \qquad \qquad \qquad
\ee
and 
\ba
\nn
S_2(AD) &=& -\log\tr\rho_{AD}^2\\
\nn
& =& - \log \Big[ \frac{1}{D^2} \Big(\frac{1}{\pi}\Big)^{8N} \int d^3p_{i'} d^3p_i d^3p_{r'} d^3p_g \int d^2\{\beta_{i'a}^{l}\} d^2\{\beta_{ia}^{l}\} d^2\{\beta_{r'a}^{l}\} d^2\{\beta_{fa}^{l}\} d^2\{\beta_{g'a}^{l}\} d^2\{\beta_{ga}^{l}\} d^2\{\beta_{h'a}^{l}\} d^2\{\beta_{la}^{l}\}\\
\nn
&& \braket{p_i, \{\beta_{ia}^{l}\} | S^{\dagger} | p_{i'}, \{\beta_{i'a}^{l}\}} \braket{\{\beta_{i'a}^{l}\} | \{\beta_{r'a}^{l}\}} \braket{p_{r'}, \{\beta_{r'a}^{l}\} | S | p_i, \{\beta_{fa}^{l}\}} \braket{\{\beta_{fa}^{l}\} | \{\beta_{ga}^{l}\}} \braket{p_g, \{\beta_{ga}^{l}\} | S^{\dagger} | p_{r'}, \{\beta_{g'a}^{l}\}}\\
&& \braket{\{\beta_{g'a}^{l}\} | \{\beta_{h'a}^{l}\}} \braket{p_{i'}, \{\beta_{h'a}^{l}\} | S | p_g, \{\beta_{la}^{l}\}} \braket{\{\beta_{la}^{l}\} | \{\beta_{ia}^{l}\}} \Big]\,;
\ea

and 

\be
S_2(D) = -\log\tr\rho_{D}^2 = -\log\tr \Big[ \frac{d_{A}^2}{D^2} \sum_{\{m'_{s'}\}} \ket{\{k_{s'}^{(\lambda_{s'})}\}_{m'_{s'}}}_{out \ out} \bra{\{k_{s'}^{(\lambda_{s'})}\}_{m'_{s'}}} \Big] = -\log \Big[ \frac{d_{A}^2d_{B}}{D^2} \Big] = \log d_{B}\,. \quad \quad \quad \qquad \qquad \quad
\ee
We can now evaluate the tripartite mutual information $I_{3(2)}$ as
\ba
\label{57}
\nn
&& I_{3(2)} = S_2(C) + S_2(D) - S_2(AC) - S_2(AD)\\
\nn
&& = \log d_{A} + \log d_{B} + \log \Big[ \frac{1}{D^2} \Big(\frac{1}{\pi}\Big)^{8N} \int d^3p_{i'} d^3p_i d^3p_{r'} d^3p_r \int d^2\{\beta_{i'a}^{l}\} d^2\{\beta_{ia}^{l}\} d^2\{\beta_{r'a}^{l}\} d^2\{\beta_{ra}^{l}\} d^2\{\beta_{g'a}^{l}\}\\
\nn
&& d^2\{\beta_{ga}^{l}\} d^2\{\beta_{h'a}^{l}\} d^2\{\beta_{ha}^{l}\} \braket{p_i, \{\beta_{ia}^{l}\} | S^{\dagger} | p_{i'}, \{\beta_{i'a}^{l}\}} \braket{\{\beta_{i'a}^{l}\} | \{\beta_{r'a}^{l}\}} \braket{p_{r'}, \{\beta_{r'a}^{l}\} | S | p_r, \{\beta_{ra}^{l}\}} \braket{\{\beta_{ra}^{l}\} | \{\beta_{ia}^{l}\}}\\
\nn
&& \braket{p_r, \{\beta_{ga}^{l}\} | S^{\dagger} | p_{r'}, \{\beta_{g'a}^{l}\}} \braket{\{\beta_{g'a}^{l}\} | \{\beta_{h'a}^{l}\}} \braket{p_{i'}, \{\beta_{h'a}^{l}\} | S | p_i, \{\beta_{ha}^{l}\}} \braket{\{\beta_{ha}^{l}\} | \{\beta_{ga}^{l}\}} \Big] + \log \Big[ \frac{1}{D^2} \Big(\frac{1}{\pi}\Big)^{8N}\\
\nn
&& \int d^3p_{i'} d^3p_i d^3p_{r'} d^3p_g \int d^2\{\beta_{i'a}^{l}\} d^2\{\beta_{ia}^{l}\} d^2\{\beta_{r'a}^{l}\} d^2\{\beta_{fa}^{l}\} d^2\{\beta_{g'a}^{l}\} d^2\{\beta_{ga}^{l}\} d^2\{\beta_{h'a}^{l}\} d^2\{\beta_{la}^{l}\} \braket{\{\beta_{la}^{l}\} | \{\beta_{ia}^{l}\}}\\
\nn
&& \braket{p_i, \{\beta_{ia}^{l}\} | S^{\dagger} | p_{i'}, \{\beta_{i'a}^{l}\}} \braket{\{\beta_{i'a}^{l}\} | \{\beta_{r'a}^{l}\}} \braket{p_{r'}, \{\beta_{r'a}^{l}\} | S | p_i, \{\beta_{fa}^{l}\}} \braket{\{\beta_{fa}^{l}\} | \{\beta_{ga}^{l}\}} \braket{p_g, \{\beta_{ga}^{l}\} | S^{\dagger} | p_{r'}, \{\beta_{g'a}^{l}\}}\\
&& \braket{\{\beta_{g'a}^{l}\} | \{\beta_{h'a}^{l}\}} \braket{p_{i'}, \{\beta_{h'a}^{l}\} | S | p_g, \{\beta_{la}^{l}\}} \Big].
\ea
This a rather convoluted expression, which nevertheless will allow us to draw relevant  consequences in the next sections.

\subsection{Tripartite mutual information at the leading-order} \label{tripa}
\noindent
The expression of the $S$ matrix generic element $\braket{p_{i'}, \{\beta_{i'a}^{l}\} | S | p_{i}, \{\beta_{ia}^{l}\}}$ in \eqref{46} has been expressed previously. We deploy now that result, substituting it into the integrand of $I_{3(2)}$, and finally expand it at the leading-order. After a long calculation,  neglecting sub-subdominant interaction terms while focusing only on the zero-th order contributions, we may derive the tripartite mutual information for the asymptotically free particles, we obtain
\ba
\nn
&& I_{3(2)}^0 = S_2(C) + S_2(D) - S_2(AC) - S_2(AD)\\
\nn
&& = \log d_{A} + \log d_{B} + \log \Big[ \frac{1}{D^2} \Big(\frac{1}{\pi}\Big)^{2N} \int d^3p_i d^3p_r \delta^{(3)}(0) \delta^{(3)}(0) \int d^2\{\beta_{ia}^{l}\} d^2\{\beta_{ga}^{l}\} \braket{\{\beta_{ia}^{l}\} | \{\beta_{ia}^{l}\}} \braket{\{\beta_{ga}^{l}\} | \{\beta_{ga}^{l}\}} \Big]\\
\nn
&& + \log \Big[ \frac{1}{D^2} \Big(\frac{1}{\pi}\Big)^{N} \int d^3p_i d^3p_g \delta^{(3)}(p_{i} - p_{g}) \delta^{(3)}(p_{g} - p_{i}) \int d^2\{\beta_{ia}^{l}\} \braket{\{\beta_{ia}^{l}\} | \{\beta_{ia}^{l}\}} \Big]\\
&& = \log d_{A} + \log d_{B} + \log \Big[ \frac{d_A^2 d_B^2}{D^2} \Big] + \log \Big[ \frac{d_A d_B}{D^2} \Big] = 0.
\ea
\end{widetext}
This approximated result  actually conforms to na\"ive expectations that the free particles should not be scrambling.
We can  build upon this exemplified paradigm taking in account interaction terms. 
Adding contributions that are provided to $S_2(AC)$ and $S_2(AD)$ at leading-order by the interaction terms -- see e.g. Appendix \ref{A-E} --- and using Eq.~\eqref{D12} -- see e.g. Appendix \ref{A-D} --- we obtain
\ba \label{pf}
\nn
I_{3(2)} &=& S_2(C) + S_2(D) - S_2(AC) - S_2(AD)\\
\nn
&=& \log d_{A} + \log d_{B} + \log \Big[ \frac{d_A^2 d_B^2}{D^2} \Big] +\log \Big[ \frac{d_A d_B}{D^2} \Big]
\nn \\
&+& 2\log \Big[ 1 
\!-\! \frac{4 \pi T}{d_A} \!\int \! d^3p_i d^3p_j  \delta(p_j^0-p_i^0) |m_{0,0} (p_{i}, p_{j})|^2 \Big] \nn \\
&=& 2 \log \Big[ 1 \!- \!\frac{4 \pi T}{d_A} \!\int \! d^3p_i d^3p_j  \delta(p_j^0-p_i^0) |m_{0,0} (p_{i}, p_{j})|^2 \Big] \nn\\
&\simeq& - \frac{8 \pi T}{d_A} \int d^3p_i d^3p_j  \delta(p_j^0-p_i^0) |m_{0,0} (p_{i}, p_{j})|^2 \nn\\
&=& - \frac{8 \pi T}{d_A} \int d|p_i| d\theta_i \ 16\pi^2 \cos(\theta_i/2) \nn \\
&\phantom{=}& \times \frac{(2 \pi \alpha Z)^2}{4 m^2 \sin^3(\theta_i/2)} \big[ |p_i|^2 \cos^2(\theta_i/2) + m^2 \big]\,,
\ea
having used that $D=d_A d_B$.\\

The final result can be the achieved specifying the integral with respect to $\theta_i$ and the integral with respect to $p_i$. Even though the integral with respect to $\theta_i$ diverges at $\theta_i=0$, we shall not be concerned by such a non physical divergence. Guidance in stead is provided by the  experimental results, which display a peak at $\theta_i=0$ that cannot be interpreted as a divergence. \\

Expanding \eqref{pf}, we can isolate two terms, and use $A_1$ and $A_2$ to label the two integrals with respect to $\theta_i$, which are positive for $\theta_i \in [0,\pi]$. Then the tripartite mutual information $I_{3(2)}$ casts
\ba
I_{3(2)} &=& - \frac{8 \pi T}{d_A} \ \frac{4\pi^2 (2 \pi \alpha Z)^2}{m^2} \ A_1 \int |p_i|^2 d|p_i| \nn\\
&\phantom{=}& - \frac{8 \pi T}{d_A} \ 4\pi^2 (2 \pi \alpha Z)^2 \ A_2 \int d|p_i|\,.
\ea

To calculate the integral with respect to $p_i$ some additional steps are required. Considering that inside a box of volume $V$ momenta are discretized as ${\bf p}= {\bf n} (2\pi/L)$, where $V=L^3$ and ${\bf n}=(n_1, n_2, n_3)$, with $n_1, n_2, n_3\in \mathbb{N}$, we can express the dimension $d_A$ of the space of hard particle states $A$ as the sum over the modes ${\bf n}$, namely    
\ba
\label{ppp}
 \int d^3p = 
\frac{(2\pi)^3}{V} \, \sum_{n} = (2\pi)^3 \frac{d_A}{V}\,,
\ea
and then immediately find a useful related equality   
\ba
\int d|p| = |p| = 2\pi \left( \frac{3 d_A}{4\pi V} \right)^{\frac{1}{3}}.
\ea

We are ready to combine all the contributions, and finally obtain as the tripartite mutual information $I_{3(2)}$, which reads
\ba
\label{i3f}
I_{3(2)} &=& - 8\pi^2 (2 \pi \alpha Z)^2 A_1 (2\pi)^3 \frac{T}{m^2 V} \nn\\
&\phantom{a}& - 64\pi^4 (2 \pi \alpha Z)^2 A_2 \big( \frac{3}{4\pi}\big)^{\frac{1}{3}} d_A^{-\frac{2}{3}} \frac{T}{V^{\frac{1}{3}}}\,.
\ea

\section{Discussion} \label{dis} 
\noindent 
Keeping the ratio between $T$ and $V$ fixed, the second term in \eqref{i3f} is negligible with respect to the first term, and can be ignored, i.e.

\ba
\label{i3fB}
I_{3(2)} \simeq - 8\pi^2 (2 \pi \alpha Z)^2 A_1 (2\pi)^3 \frac{T}{m^2 V} \,.
\ea
The dimension $d_A$ of the Hilbert space endowed with a proper regularization is finite. Without a regularization procedure, the dimension $d_A$ of the Hilbert space of the states limited within a box would be infinite.
The volume $V$ of the box, and the time interval $T$ of the scattering processes are finite quantities. Their ratio $T/(m^2V)$, which is dimensionless in natural units, at fixed values of $T$ and $V$ or for a proper relative scaling is easily shown to be also finite, and to be much smaller than $2\log{d_A}$. Once a proper regularization is adopted, both $d_A$ and $\int d^3p$ turn out to be finite, suggesting that the relation between $d_A$ and $V$ will not change. Similarly, the relation between $d_A$ and $T/V$ will not be altered by the regularization procedure, ensuring the final result to hold consistently before and after regularization. This finally implies that $I_{3(2)}$ will be much larger than the lower bound of the range of tripartite mutual information. \\

The tripartite mutual information is inverse proportional to the mass of the hard particles, i.e. to the mass of the electron in our case. In the infinite mass limit, hard particle decouple from the other subsystems, and the contribution to the tripartite mutual information becomes vanishing. In other words, no scrambling of information is encountered in this limit, and the subsystems retain the information stored by the coherence of their degrees of freedom. \\

The tripartite mutual information, both in the approximated form in Eq.~\eqref{i3fB} and in its generic form in Eq.~\eqref{i3f}, is vanishing in the large $T$ and $V$ limit if $T/V\rightarrow{0}$. This behavior has a simple interpretation: when the space dimensions of the scattering center exceed the scaling of the time dimension, causality suppress decoherence, and hence scrambling among the subsystems constituents cannot happen. Conversely, if we reshuffle the relevant result in Eq.~\eqref{i3fB} in terms of the mass density $\rho\sim m/V$ and the energy density resolution $m^3/T$, an isotropic scaling at fixed  $Z$ --- number of electrons of the scattering center ---  among these quantities, which appear in a ratio within Eq.~\eqref{i3fB}, solely relates scrambling to the second power of the fine-structure constant $\alpha$. This means that for vanishing electromagnetic interaction, when no interaction with the environment are considered, namely in the $\alpha\rightarrow{0}$ limit, no decoherence can take place, and accordingly no scrambling of information can be attained. \\

Consequences that can be driven out of this analysis are profound, in that our result urges to reconsider the black hole information paradox \cite{Hawking1,Hawking2}. In perturbative QED, asymptotic states are deployed while addressing a description of scattering processes. Scattering takes place at very short distances, which are dealt with, within the effective field theory approach, as points of zero measure. Nonetheless, physically scattering processes, as any physical processes in general, must happen in a finite space region. Deploying an energy $E$ that is the one in the center of mass of the scattering particles, we can resolve distances $L$, through scattering processes that localize a minimum amount of energy $E>1/L$ within a space-time hypercube of edge $L$. The Schwarzschild radius associated to this localized system is $R_{\rm S}=L_P^2/L$, with $L_P=\sqrt{G}$ denoting the Planck length in natural units. At smaller distances than the Planck length, namely for  $L <\!\!<L_P$, the Schwarzschild radius of the system exceeds both $L_P$ and $L$. Thus the energy of the system shall be localized  within a distance that is much smaller than $R_{\rm S}$, which is clearly impossible \cite{Dvali:2010bf}. A classical macroscopic black hole hence emerges in this measurement procedure associated to a generic scattering process. Being $L<\!\!<L_P$, there is no information can be extracted within the horizon of this black hole, with macroscopic radius $R_{\rm S}=L_P^2/L>\!\!>L_P>\!\!>L$, consistently with the holographic principle picture \cite{Hol1, Hol2, Hol3}. \\

Even though a black hole can thought to be effectively present through this measurement procedure, ordinary quantum field theoretical tools, as well as a consolidated experimental guidance provided by a wealth of particle physics data, ensure us that anyway scattering processes takes place, and more importantly that the asymptotic states can be interpreted as the byproduct of a Hawking radiation\footnote{Modelling an effective formation/presence of a black hole is just an artifact of the point measurement assumption, which looses its meaning in a quantum theory. That the scattering process actually takes place tells us that this sub-$L_P$ scale assumption has no physical validity --- otherwise, the LHC would have collapsed into a black hole.}. This approach to Hawking radiation is exactly in the spirit of the seminal paper \cite{Hawking1}. \\

This simple heuristic approach, which nevertheless is well grounded on the physical concepts that are at the base of black hole holography and black hole radiation, enables to extend our results to the framework of the black hole information paradox. The picture we have been drawing share similarities with the idea that black holes are Bose-Einstein condensates of gravitons, as propose by Dvali and Gomez in Ref.~\cite{Dvali:2011aa,Dvali:2012en,Dvali:2012wq}. Nonetheless, our results do not necessarily rely on this interpretation, and are more general.\\

The specificity of our approach can be rather summarized in light of its consequences. First, we have seen that scrambling happens even though the $S$ matrix elements that have been recovered are fully unitary. So, scrambling of information does not either require or imply loss of unitarity. Second, decoherence due to the interaction with the environment is essential to realize the scrambling process. This happens precisely because of the emission-absorption of infrared radiation: soft photons remove information from the incoming hard state and transfer to the environment, where it is observationally inaccessible and hence ``lost'' as far as the measured hard-state outcomes are concerned. This transfer of information, which is the usual kind of environment-induced decoherence, represents the main difference among our and previous results in the literature. Our procedure, by adopting a tripartite mutual information scheme, allows to introduce in the picture the role of the environment, and enables to consider decoherence through the interaction modeled by means of QFT ---  in particular QED in our analysis. In other words, what we have calculated here is how much information is lost to the environment.\\

It is also worth mentioning that soft radiation represents in this scenario a class of universality that encode several options considered in the literature while resorting to hidden symmetries and investigating their role for the solution of the black hole information paradox. Notably, the possibility of using BMS symmetries --- see e.g. Ref.~\cite{Bern:2014oka} for a quantum perturbative treatment ---  was proposed by Hawking, Perry and Strominger in \cite{Hawking:2016msc}. Kac-Moody symmetries \cite{KacMoody1,KacMoody2,KacMoody3} have been also advocated to achieve an encoding of information \cite{Addazi:2017xur} through the hidden symmetries charges in order to solve the black information loss paradox. The novelty of our approach stands in recognizing that is the interaction with the soft bosons emission and absorption, which naturally require to give a role to the environment, is responsible for the scrambling power of the scattering process analogous to the Hawking radiation effect.\\     

Summarizing, in our picture the black hole information loss paradox is solved, while evolving asymptotic states of QFT with unitary operators, because of a decohering interaction with the environment that is mediated by the soft bosons.

\section{Conclusions and outlooks}\label{conclu}
\noindent
We have addressed the issue of scrambling in quantum information processing, focusing on the paradigmatic study-case of the scattering of an electron by an electromagnetic source in QED that involves emission-absorption of soft photons. In order to achieve this result, we calculated the tripartite mutual information of the process.\\

Since it is highly nontrivial to attain a diagonalization of all the reduced density matrices that are involved in the specific case we have here considered, in stead of calculating the tripartite mutual information in terms of Von Neumann entropy, we have resorted to the calculation of the tripartite mutual information in terms of the 2-Renyi entropy. We then found that the tripartite mutual information hence estimated is negative, finite and much larger than the lower bound $-2 \log d_A$. On the other hand, the tripartite mutual information obtained in terms  the 2-Renyi entropy provides an upper bound for the tripartite mutual information calculated in terms of von Neumann entropy, which finally enables us to interpret our results as bounds for this latter quantity.\\

Finally, we can state that the scattering process provided with soft photons emission-absorption is scrambling, and that the scrambling power of the process entirely relies on the (QED) subsystems' interaction. As a byproduct of this analysis, we find there exists an amount of information that is indeed encoded in the soft photons, and that the information carried out by the hard part of the incoming states cannot be obtained {\it sic et simpliciter} by local measurements of the outgoing states, as it should be actually  greater than, or equal to, the absolute value of what the information we have calculated. Although for the specific process we have analyzed we still expect the amount of scrambled information to be small, having obtained the estimate of this information by means of local measurements, still this is sufficient to the relevance of our statement. \\

Our procedure provides a new look at the black hole information loss paradox. Indeed, in our approach scrambling is attained by decoherence of the scattering system hard constituents. On the other hand, this decoherence can only happen because of an (electromagnetic) interaction with the environment is present, due to the emission-absorption of soft photons. Because of previous attempts questioning preservation of unitarity in the Hawking radiation effect, it is relevant to emphasize that the whole process has been described only resorting to a fully unitary evolution of the scattering system.\\

We argue that our result, which has been derived in the specific framework of QED, still is generic enough to be hold generically in QFT and account for general thermally correlated states\cite{chirco}. Curvature can be easily accounted for in this framework, considering interaction with a Bose-Einstein condensate of gravitons \cite{SHM}, along the line of \cite{Dvali:2011aa,Dvali:2012en,Dvali:2012wq}, providing an extension of our analysis so as to account for the scrambling power of soft gravitons \cite{JHEP2013159}. Although we expect that qualitatively the outcome will not change, this will clarify in greater generality the scrambling power of gauge interactions in nature. 

\section*{Acknowledgments}
\noindent
A.M. wished to thank Chris Fields and Matteo Lulli for enlightening discussions.  
A.M. wishes to acknowledge support by the NSFC, through the grant No. 11875113, the Shanghai Municipality, through the grant No.~KBH1512299, and by Fudan University, through the grant No.~JJH1512105. A.H. acknowledges support from NSF award no. 2014000.

\appendix

\section{Properties of the harmonic oscillator coherent state}
\noindent
We review here some basic properties of the harmonic oscillator coherent states, in order to provide a solid basis for the calculations reported in the main body of this paper. The coherent states of the harmonic oscillator, which we denote with $\ket{\alpha}$, are the eigenstates of the annihilation operator $a$, namely 
\be
a\ket{\alpha} = \alpha\ket{\alpha}\,,
\ee
with $\alpha$ complex eigenvalue.
A coherent state $\ket{\alpha}$ is generated from the vacuum $\ket{0}$ by the action of a displacement operator $D(\alpha)$
\be
\ket{\alpha} = D(\alpha)\ket{0} = \exp[\alpha a^{\dagger} - \alpha^* a]\ket{0}\,.
\ee
By means of a formal expansion of the exponential, we can find the expression of the coherent states in terms of the Fock states
\ba
\nn
\ket{\alpha} &=& \exp\Big(-\frac{1}{2}|\alpha|^2\Big) \sum_{n} \frac{(\alpha a^{\dagger})^n}{n!} \ket{0}\\
&=& \exp\Big(-\frac{1}{2}|\alpha|^2\Big) \sum_{n} \frac{\alpha^n}{\sqrt{n!}} \ket{n}\,.
\ea
The scalar product among two coherent states reads
\ba
&& \braket{\beta|\alpha} = \exp\Big( -\frac{1}{2}|\alpha|^2 -\frac{1}{2}|\beta|^2 + \alpha\beta^* \Big)\,,\\
&& |\braket{\beta|\alpha}|^2 = \exp\Big(-|\alpha-\beta|^2\Big)\,.
\ea
Coherent states are not orthogonal among one another, since the scalar product of two coherent states $\ket{\alpha}$ and $\ket{\beta}$ does not vanish. Nonetheless, coherent states represent a span of the Fock space, i.e.
\ba
\nn
&& \frac{1}{\pi} \int d^2\alpha \ \ket{\alpha} \bra{\alpha}\\
\nn
&=& \frac{1}{\pi} \int d^2\alpha \exp(-|\alpha|^2)\sum_{mn}\frac{\alpha^m\alpha^{*n}}{\sqrt{m!n!}} \ket{m} \bra{n}\\
\nn
&=& \frac{1}{\pi} \sum_{mn} \frac{1}{\sqrt{m!n!}} \int rdr \e^{-r^2} r^{m+n} \int d\phi \e^{i(m-n)\phi} \ket{m} \bra{n}\\
&=& \frac{1}{\pi} \sum_{mn} \frac{1}{\sqrt{m!n!}} \pi m! \delta_{mn} \ket{m} \bra{n} = \sum_n \ket{n} \bra{n} = \mathbb{I}\,.
\ea
The volume of the coherent state can be also calculated
\ba
\nn
&& \int d^2 \alpha \braket{\alpha | \alpha}\\
\nn
&=& \int d^2 \alpha \exp(-|\alpha|^2) \sum_{nm} \frac{(\alpha)^m(\alpha^*)^n}{\sqrt{m!n!}} \braket{n | m}\\
\nn
&=& \int d^2 \alpha \exp(-|\alpha|^2) \sum_{n} \frac{|\alpha|^{2n}}{n!}\\
&=& \sum_{n} \frac{1}{n!} \int rdr \e^{-r^2}r^{2n} \int d\phi = \sum_{n} \pi = n \pi = \pi d\,, \quad \
\ea
having introduced in the last equality the dimension of the Fock-space states at some fixed momentum and polarization, namely $d=\sum_{n}$.\\

A notable integral relevant to our calculations is
\ba
\nn
&& \int d^2 \alpha d^2 \beta |\braket{\beta|\alpha}|^2 = \int d^2 \alpha d^2 \beta \exp (-|\alpha - \beta|^2)\\
&& = \int d^2 \alpha d^2 \beta \braket{\beta|\alpha} \braket{\alpha|\beta} = \pi \int d^2 \alpha \braket{\alpha|\alpha} = \pi^2 d\,. \quad
\ea

A generalized coherent state can also be defined as the direct product of several copies of differently labelled coherent states

\ba \label{gcs}
\nn
\ket{\{\alpha_i\}} &=& \prod_{i}\ket{\alpha_i} = D(\{\alpha_i\})\ket{0}\\
&=& \exp[\sum_i \alpha a_i^{\dagger} - \alpha_i^* a_i]\ket{0}\,.
\ea

For $i=1, \dots N$, the relevant completeness relation is provided by the expression

\be
\Big(\frac{1}{\pi}\Big)^N \int d^2\{\alpha_i\} \ \ket{\{\alpha_i\}} \bra{\{\alpha_i\}} = \mathbb{I}\,. 
\ee

The volume of a generalized coherent state, as it appears in \eqref{gcs}, is 

\ba
\nn
\int d^2 \{\alpha_i\} \braket{\{\alpha_i\} | \{\alpha_i\}} &=& \prod_i \int d^2 \alpha_i \braket{\alpha_i | \alpha_i}\\
&=& \prod_i \pi d_i = \pi^N d\,.
\ea

Another integral that has been useful for our calculations is

\be
\int d^2 \{\alpha_i\} d^2 \{\beta_i\} |\braket{\{\beta_i\} | \{\alpha_i\}}|^2 = \pi^{2N} d\,.
\ee

Further details on properties of the harmonic oscillator coherent states can be found in several textbooks --- see also \cite{PRD94123517} for a short review.

\section{Fock states and Coherent states}
\noindent
Usually, the photon Fock state is expressed as $\ket{\{k_i^{(\lambda_i)}\}_m}$, with $i = 1 \cdots m$. This expression denote the presence of $m$ photons, with momenta $k_{i}$ and polarizations $\lambda_i$. Since photons are bosons, an unlimited number of photons can be found in the same state. So, equivalently, we can express the  photon Fock state as $\ket{\{k_j^{(\lambda_i)}\}_{n_{j}}}$, with $\sum_{j} n_{j} = m$. This denotes the presence of $m$ photons, $n_{j}$ of  which with momentum $k_{j}$ and polarization $\lambda_j$. Since at fixed momentum and polarization, a Fock-space state is also an eigenstate of the number operator, we can find a one-to-one correspondence among the states
\ba
\ket{\{k_j^{(\lambda_i)}\}_{n_{j}}} \Longrightarrow \ket{n_{j}(k_{j}^{(\lambda_j)})}\,.
\ea
We recall the relevance of these latter states in connecting Fock states to coherent states.\\

\begin{figure}[h]
\centering
\includegraphics[width=8cm,height=4.8cm]{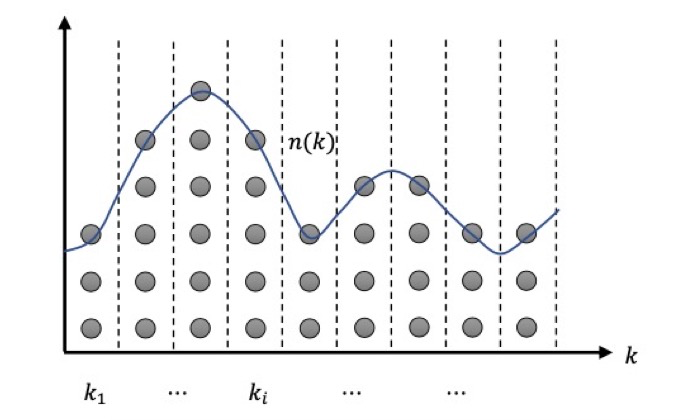}
\caption{Schematic diagram of the states of the photon Fock-space.}
\label{figure3}
\end{figure}

The definition of the photon coherent state is
\ba
\nn
\ket{\beta_{a}^l} &=& D(\beta_{a}^l) \ket{0} = \text{exp}\Big(-\frac{1}{2} |\beta_{a}^l|^2\Big) \text{exp}\Big(\beta_{a}^l a_{a}^{l{\dagger}}\Big) \ket{0}\\
\nn
&=& \text{exp}\Big(-\frac{1}{2} |\beta_{a}^l|^2\Big) \sum_{n} \frac{(\beta_{a}^l)^n}{\sqrt{n!}} \ket{n_{l,a}}\,.
\ea
The states denoted by $\ket{n_{l,a}}$ are the relevant eigenstates of the number operator. In other words, there are $n_{l,a}$ photons with momentum distribution $f_{a}(k)$ and polarization $l$. So there must be $\sum_{a} n_{l,a} f_{a}(k_j) = n_{j}(k_{j}^{(l)})$ photons with momentum $k_{j}$ and polarization $l$. Thus, there exists a one-to-one correspondence among the states:
\ba
\ket{n_{j}(k_{j}^{(l)}} \Longleftrightarrow \ket{\sum_{a} n_{l,a} f_{a}(k_j)}.
\ea

The one-to-one correspondence among the states of the photon Fock-space and the photon coherent states implies that in our calculation we can safely transform states of the photon Fock-space into photon coherent states. The dimensions of the two spaces are the same, states can be expanded in either bases without any ambiguity, and their inner product is always well defined.

\begin{figure}[h]
\centering
\includegraphics[width=8cm,height=4.5cm]{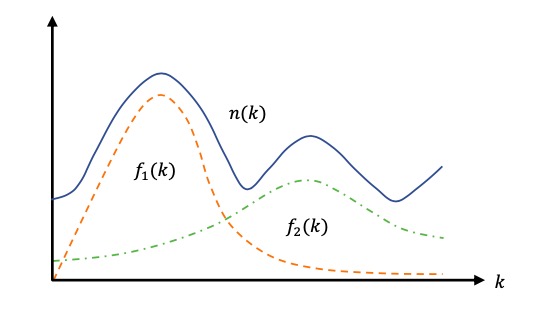}
\caption{Schematic diagram of photon coherent states.}
\label{figure4}
\end{figure}

\section{Continuous states and box states}\label{A-C}
\noindent
Continuous states $\psi_{{p_1},{\sigma_1},{n_1};{p_2},{\sigma_2},{n_2};\cdots}$ --- where $p$ label momentum, $\sigma$ label helicity and $n$ label particle type --- are normalized according to the following formula \cite{QFTSW}
\ba
\label{refC1}
\nn
&& \braket{\psi_{{p'_1},{\sigma'_1},{n'_1};{p'_2},{\sigma'_2},{n'_2};\cdots}|\psi_{{p_1},{\sigma_1},{n_1};{p_2},{\sigma_2},{n_2};\cdots}}\\
\nn
&=& \delta^{(3)}(\boldsymbol{p}'_1 - \boldsymbol{p}_1)\delta_{\sigma'_1,\sigma_1}\delta_{n'_1,n_1}\delta^{(3)}(\boldsymbol{p}'_2 -\boldsymbol{p}_2)\delta_{\sigma'_2,\sigma_2}\delta_{n'_2,n_2}\cdots\\
&& \pm \text{permutations}\,,
\ea
where $\delta_{\sigma',\sigma}$ and $\delta_{n',n}$ denotes the ordinary Kronecker delta. The odd permutation of half-integer spin particles entails the sign $-1$, otherwise the sign is $+1$. It is usually convenient to employ a short notation and denote the collection of set of indices $\{\{p_1\},\{\sigma_1\},\{n_1\};\{p_2\},\{\sigma_2\},\{n_2\};\cdots\}$ with
a Greek letter, e.g. $\alpha$. Within this notation, the normalization condition \eqref{refC1} recasts
\be
\braket{\psi_{\alpha'}|\psi_{\alpha}} = \delta(\alpha'-\alpha)\,.
\ee
Also, while summing over the states, we write

\be
\int d\alpha \cdots = \sum_{n_1\sigma_1n_2\sigma_2} \int d^3p_1 d^3p_2 \cdots\,.
\ee
Constraining the whole system of physical particles (continuous states) in a large box with macroscopic volume $V$, one obtains a discrete set of box states. Being spatial wave-functions single-valued, momenta are required to be quantized
\ba
\boldsymbol{p} = \frac{2\pi}{L} (n_1, n_2, n_3)\,,
\ea
where $n_i$ are integers, and the macroscopic volume $V=L^3$. Correspondingly, the three-dimensional delta functions become
\be
\delta^{(3)}_{V}(\boldsymbol{p}' - \boldsymbol{p}) = \frac{1}{(2\pi)^3} \int_V d^3x \ \e^{i(\boldsymbol{p}' - \boldsymbol{p})\cdot\boldsymbol{x}} = \frac{V}{(2\pi)^3} \delta_{\boldsymbol{p}' \boldsymbol{p}}\,,
\ee
where $\delta_{\boldsymbol{p}' \boldsymbol{p}}$ is the ordinary Kronecker delta. Due to the normalization condition \eqref{refC1}, the scalar product among the continuous states constrained within a box will turn out to be not only a sum of products of Kronecker deltas, but will also encode a factor $[V/(2\pi)^3]^N$, with $N$ the number of particles within the Fock-space state under scrutiny. As a consequence, the norm of continuous states in a box, say $\psi_\alpha$, is
expressed by 
\be
\braket{\psi_{\alpha}|\psi_{\alpha}} = \delta_{V}(0) = \big[{V}/{(2\pi)^3}\big]^{N_{\alpha}}\,.
\ee
So, normalized states in a box, or box states, read
\be
\psi_\alpha^{\text{Box}} = \big[(2\pi)^3/V\big]^{N_{\alpha}/2} \psi_{\alpha}\,,
\ee
with normalization condition
\be
\braket{\psi_\beta^{\text{Box}}|\psi_\alpha^{\text{Box}}} = \delta_{\beta\alpha}\,,
\ee
where $\delta_{\beta\alpha}$ is the Kronecker delta.

The relation between the $S$ matrix corresponding to the continuous states in the box and the one corresponding to the box states is
\be
S_{\beta\alpha} = \big[V/(2\pi)^3\big]^{(N_{\beta}+N_{\alpha})/2} S_{\beta\alpha}^{\text{Box}}\,.
\ee

Notice that, while summing over the states, the number of one-particle states in a volume $d^3p$ equals $Vd^3p/(2\pi)^3$ which is the number of integers $n_1,n_2,n_3$ in $d^3p$. This implies that

\be
\sum_{p} \cdots = \big[V/(2\pi)^3\big] \int d^3p \cdots\,.
\ee

Correspondingly, we have
\be
\sum_{\alpha} \cdots = \big[V/(2\pi)^3\big]^{N_{\alpha}} \int d\alpha \cdots\,.
\ee
It is also necessary to regulate a physical system over the time axis, constraining it an interval. This entails switching on the interaction only for a limited time $T$. The corresponding delta function imposing energy conservation becomes
\be
\delta_{T}(p'^0-p^0) = \frac{1}{2\pi}\int_{-T/2}^{T/2} dt \ \e^{i(p'^0-p^0)t} = \frac{T}{2\pi} \delta_{p'^0 p^0}\,,
\ee
where $\delta_{p'^0 p^0}$ is the Kronecker delta.

In perturbation theory, the $S$ matrix can be expressed as
\be
\label{C13}
S_{\beta\alpha} = \delta^{(3)} (p_{\beta}-p_{\alpha}) - 2i\pi\delta^{(4)}(p_{\beta} - p_{\alpha}) M_{\beta\alpha}\,, \quad
\ee
where the matrix element $M_{\beta\alpha}$ is the delta-function free contribution. In the box, the latter relation recasts
\be
S_{\beta\alpha} = \delta^{(3)}_{V}(p_{\beta}-p_{\alpha}) - 2i\pi\delta_{T}(p_{\beta}^0-p_{\alpha}^0) \delta^{(3)}_{V}(\boldsymbol{p}_{\beta} - \boldsymbol{p}_{\alpha}) M_{\beta\alpha}\,.
\ee
So the squares of delta functions in $|S_{\beta\alpha}|^2$ can be dealt with as
\ba
\nn
&\big[\delta^{(3)}_{V}(\boldsymbol{p}_{\beta} - \boldsymbol{p}_{\alpha})\big]^2& = \delta^{(3)}_{V}(\boldsymbol{p}_{\beta} - \boldsymbol{p}_{\alpha}) \delta^{(3)}_{V}(0)\\
&& = \frac{V}{(2\pi)^3} \delta^{(3)}_{V}(\boldsymbol{p}_{\beta} - \boldsymbol{p}_{\alpha})\,,
\ea

\ba
\nn
&\big[\delta_{T}(p_{\beta}^0-p_{\alpha}^0)\big]^2& = \delta_{T}(p_{\beta}^0-p_{\alpha}^0) \delta_{T}(0)\\
&& = \frac{T}{2\pi} \delta_{T}(p_{\beta}^0-p_{\alpha}^0)\,.
\ea

\section{Implications of unitarity}\label{A-D}
\noindent
The standard textbook formulation of the unitarity condition for the $S$ matrix reads
\be
\int d\beta \ S_{\beta\gamma}^* S_{\beta\alpha} = \delta^{(3)} (p_{\gamma}-p_{\alpha})\,.
\ee
Plugging in the expression of the $S$ matrix in \eqref{C13}, and simplifying the term $\delta^{(3)}(p_{\gamma}-p_{\alpha})$ and a factor $2\pi \delta^{(4)}(p_{\gamma}-p_{\alpha})$, one can obtain \cite{QFTSW}
\be
- i M_{\gamma\alpha} + i M_{\alpha\gamma}^* = - 2\pi \int d\beta \ \delta^{(4)}(p_{\beta} - p_{\alpha}) M_{\beta\gamma}^* M_{\beta\alpha}\,.
\ee
In the special case $\alpha = \gamma$, a most useful relation can be recovered 
\ba
\text{Im} \ M_{\alpha\alpha} = - \pi \int d\beta \ \delta^{(4)}(p_{\beta} - p_{\alpha}) |M_{\beta\alpha}|^2\,.
\ea
This is the unitarity condition for the states that belong to the space $\mathcal{H}_{FQ} \otimes \mathcal{H}_{F\gamma}$. Similarly, we can derive the unitarity condition for the states of the space $\mathcal{H}_{FQ} \otimes \mathcal{H}_{C\gamma}$. In this latter case, the expression of the $S$ matrix is
\begin{widetext}
\be
\label{D4}
\braket{p_{i'}, \{\beta_{i'a}^{l}\} | S | p_{i}, \{\beta_{ia}^{l}\}} = \delta^{(3)}(p_{i'} - p_i) \braket{\{\beta_{i'a}^{l}\} | \{\beta_{ia}^{l}\}} - 2i\pi \delta^{(4)}(p_{i'} - p_i) \braket{p_{i'}, \{\beta_{i'a}^{l}\} | M | p_{i}, \{\beta_{ia}^{l}\}}\,.
\ee
The unitarity condition can be then expressed as
\ba \label{aaas}
\nn
&& \delta^{(3)} (p_{j}-p_{i}) \braket{\{\beta_{ja}^{l}\} | \{\beta_{ia}^{l}\}} = \Big(\frac{1}{\pi}\Big)^{N} \int d^3p_{i'} \int d^2\{\beta_{i'a}^{l}\} \braket{p_{j}, \{\beta_{ja}^{l}\} | S^+ | p_{i'}, \{\beta_{i'a}^{l}\}} \braket{p_{i'}, \{\beta_{i'a}^{l}\} | S | p_{i}, \{\beta_{ia}^{l}\}}\\
\nn
&& = \delta^{(3)}(p_{j} - p_i) \braket{\{\beta_{ja}^{l}\} | \{\beta_{ia}^{l}\}} - 2i\pi \delta^{(4)}(p_{j} - p_i) \braket{p_{j}, \{\beta_{ja}^{l}\} | M | p_{i}, \{\beta_{ia}^{l}\}} + 2i\pi \delta^{(4)}(p_{i} - p_j) \braket{p_{j}, \{\beta_{ja}^{l}\} | M^+ | p_{i}, \{\beta_{ia}^{l}\}}\\
&& + \ 4\pi^2 \Big(\frac{1}{\pi}\Big)^{N} \int d^3p_{i'} \int d^2\{\beta_{i'a}^{l}\} \delta^{(4)}(p_{i'} - p_j) \braket{p_{j}, \{\beta_{ja}^{l}\} | M^+ | p_{i'}, \{\beta_{i'a}^{l}\}} \delta^{(4)}(p_{i'} - p_i) \braket{p_{i'}, \{\beta_{i'a}^{l}\} | M | p_{i}, \{\beta_{ia}^{l}\}}\,.
\ea
If we plug into \eqref{aaas} the expression of the $S$ matrix in \eqref{D4}, and simplify the term $\delta^{(3)}(p_{j} - p_{i}) \braket{\{\beta_{ja}^{l}\} | \{\beta_{ia}^{l}\}}$ and a factor $2\pi \delta^{(4)}(p_{j}-p_{i})$, we obtain
\ba
\nn
&& - i \braket{p_{j}, \{\beta_{ja}^{l}\} | M | p_{i}, \{\beta_{ia}^{l}\}} + i \braket{p_{j}, \{\beta_{ja}^{l}\} | M^+ | p_{i}, \{\beta_{ia}^{l}\}}\\
&& = - 2\pi \Big(\frac{1}{\pi}\Big)^{N} \int d^3p_{i'} \int d^2\{\beta_{i'a}^{l}\} \delta^{(4)}(p_{i'} - p_i) \braket{p_{j}, \{\beta_{ja}^{l}\} | M^+ | p_{i'}, \{\beta_{i'a}^{l}\}} \braket{p_{i'}, \{\beta_{i'a}^{l}\} | M | p_{i}, \{\beta_{ia}^{l}\}}\,.
\ea
For the special case $i=j$, the previous relation implies 
\ba
\nn
&& 2 \ \text{Im} \ \braket{p_{i}, \{\beta_{ia}^{l}\} | M | p_{i}, \{\beta_{ia}^{l}\}}\\
&& = - 2\pi \Big(\frac{1}{\pi}\Big)^{N} \int d^3p_{i'} \int d^2\{\beta_{i'a}^{l}\} \delta^{(4)}(p_{i'} - p_i) \braket{p_{i}, \{\beta_{ia}^{l}\} | M^+ | p_{i'}, \{\beta_{i'a}^{l}\}} \braket{p_{i'}, \{\beta_{i'a}^{l}\} | M | p_{i}, \{\beta_{ia}^{l}\}}\,.
\ea
At the leading-order (in the perturbative expansion) of the scattering process involving an electron or a positron scattered by an external field, \eqref{D4} can be written as
\be
\label{D8}
\braket{p_{i'}, \{\beta_{i'a}^{l}\} | S | p_{i}, \{\beta_{ia}^{l}\}} \sim \delta^{(3)}(p_{i'} - p_i) \braket{\{\beta_{i'a}^{l}\} | \{\beta_{ia}^{l}\}} - 2i\pi \delta(p_{i'}^0 - p_i^0) \braket{p_{i'}, \{\beta_{i'a}^{l}\} | M_e | p_{i}, \{\beta_{ia}^{l}\}}.
\ee
The unitarity condition can be then expressed as
\ba
\nn
&& \delta^{(3)} (p_{j}-p_{i}) \braket{\{\beta_{ja}^{l}\} | \{\beta_{ia}^{l}\}} = \Big(\frac{1}{\pi}\Big)^{N} \int d^3p_{i'} \int d^2\{\beta_{i'a}^{l}\} \braket{p_{j}, \{\beta_{ja}^{l}\} | S^+ | p_{i'}, \{\beta_{i'a}^{l}\}} \braket{p_{i'}, \{\beta_{i'a}^{l}\} | S | p_{i}, \{\beta_{ia}^{l}\}}\\
\nn
&& = \delta^{(3)}(p_{j} - p_i) \braket{\{\beta_{ja}^{l}\} | \{\beta_{ia}^{l}\}} - 2i\pi \delta(p_{j}^0 - p_i^0) \braket{p_{j}, \{\beta_{ja}^{l}\} | M_e | p_{i}, \{\beta_{ia}^{l}\}} + 2i\pi \delta(p_{i}^0 - p_j^0) \braket{p_{j}, \{\beta_{ja}^{l}\} | M_e^+ | p_{i}, \{\beta_{ia}^{l}\}}\\
&& + 4\pi^2 \Big(\frac{1}{\pi}\Big)^{N} \int d^3p_{i'} \int d^2\{\beta_{i'a}^{l}\} \delta(p_{i'}^0 - p_j^0) \braket{p_{j}, \{\beta_{ja}^{l}\} | M_e^+ | p_{i'}, \{\beta_{i'a}^{l}\}} \delta(p_{i'}^0 - p_i^0) \braket{p_{i'}, \{\beta_{i'a}^{l}\} | M_e | p_{i}, \{\beta_{ia}^{l}\}}\,.
\ea
Plug into \eqref{D8} the expression of the $S$ matrix, and simplifying the term $\delta^{(3)}(p_{j} - p_{i}) \braket{\{\beta_{ja}^{l}\} | \{\beta_{ia}^{l}\}}$ and a factor $2\pi \delta^{(4)}(p_{j}^0-p_{i}^0)$, we obtain
\ba
\nn
&& - i \braket{p_{j}, \{\beta_{ja}^{l}\} | M_e | p_{i}, \{\beta_{ia}^{l}\}} + i \braket{p_{j}, \{\beta_{ja}^{l}\} | M_e^+ | p_{i}, \{\beta_{ia}^{l}\}}\\
&& = - 2\pi \Big(\frac{1}{\pi}\Big)^{N} \int d^3p_{i'} \int d^2\{\beta_{i'a}^{l}\} \delta(p_{i'}^0 - p_i^0) \braket{p_{j}, \{\beta_{ja}^{l}\} | M_e^+ | p_{i'}, \{\beta_{i'a}^{l}\}} \braket{p_{i'}, \{\beta_{i'a}^{l}\} | M_e | p_{i}, \{\beta_{ia}^{l}\}}\,.
\ea
once again, focusing on the special case $i=j$, we find
\ba
\nn
&& 2 \ \text{Im} \ \braket{p_{i}, \{\beta_{ia}^{l}\} | M_e | p_{i}, \{\beta_{ia}^{l}\}}\\
&& = - 2\pi \Big(\frac{1}{\pi}\Big)^{N} \int d^3p_{i'} \int d^2\{\beta_{i'a}^{l}\} \delta(p_{i'}^0 - p_i^0) \braket{p_{i}, \{\beta_{ia}^{l}\} | M_e^+ | p_{i'}, \{\beta_{i'a}^{l}\}} \braket{p_{i'}, \{\beta_{i'a}^{l}\} | M_e | p_{i}, \{\beta_{ia}^{l}\}}\,.
\ea
If we now substitute in this latter expression the $M$ matrix element found in \eqref{47}, we finally derive the expression
\ba
\label{D12}
\nn
&& 2 \ \text{Im} \ \braket{p_{i}, \{\beta_{ia}^{l}\} | M_e | p_{i}, \{\beta_{ia}^{l}\}} = - 2\pi \Big(\frac{1}{\pi}\Big)^{N} \int d^3p_{i'} \int d^2\{\beta_{i'a}^{l}\} \delta(p_{i'}^0 - p_i^0) \exp (2\alpha B (p_i, p_{i'}) + 2\alpha \tilde{B} (p_i, p_{i'}))\\
\nn
&& \exp \Big\{ - \sum_{l,a} |\epsilon_{i'a}^{l} -  \epsilon_{ia}^{l}|^2 \Big\} \Big\{ \sum_{m,m'=0}^{\infty} m_{m,m'} (p_{i}, p_{i'}) \Big\} \Big\{ \sum_{m,m'=0}^{\infty} m_{m,m'} (p_{i}, p_{i'}) \Big\}^*\\
\nn
&& = - 2\pi \Big(\frac{1}{\pi}\Big)^{N} \int d^3p_{i'} \int d^2\{\beta_{i'a}^{l}\} \delta(p_{i'}^0 - p_i^0) \big( 1+ 2\alpha B (p_i, p_{i'}) + 2\alpha \tilde{B} (p_i, p_{i'}) + o(\alpha) \big) \exp \Big\{ - \sum_{l,a} |\epsilon_{i'a}^{l} -  \epsilon_{ia}^{l}|^2 \Big\}\\
&& \Big\{ |m_{0,0} (p_{i}, p_{i'})|^2 + o(\alpha^2) \Big\}\,.
\ea

\section{Calculation of reduced density matrices}\label{A-E}
In section \ref{unitary}, we construct the matrix expression of the unitary operator $U$ in terms of the states in $\mathcal{H}_{FQ} \otimes \mathcal{H}_{C\gamma}$, more precisely in terms of the outgoing state in $\mathcal{H}_{FQ} \otimes \mathcal{H}_{C\gamma}$. Then relevant expression of the Choi states is
\ba
\nn
&& \ket{U} = (I_{in} \otimes U_{out}) \ket{I} = \frac{1}{\sqrt{D}} \sum_{\{n_{j'}\}} \int d^3p_{i'} \Big( \ket{p_{i'}, \{k_{j'}^{(\lambda_j')}\}_{n_{j'}}} \Big) \otimes U \Big( U \ket{p_{i'}, \{k_{j'}^{(\lambda_j')}\}_{n_{j'}}} \Big)\\
\nn
&& = \frac{1}{\sqrt{D}} \sum_{\{n_{j'}\}} \int d^3p_{i'} \Big[ \Big(\frac{1}{\pi}\Big)^N \int d^3p_{g} \int d^2\{\beta_{ga}^{l}\} \braket{\{\beta_{ga}^{l}\} | \{k_{j'}^{(\lambda_j')}\}_{n_{j'}}} \delta^{(3)}(p_{g} - p_{i'}) \ \ket{p_{g}, \{\beta_{ga}^{l}\}} \Big] \otimes \Big[ \Big(\frac{1}{\pi}\Big)^{3N} \int d^3p_i d^3p_l\\
\nn
&& \int d^2\{\beta_{ka}^{l}\} d^2\{\beta_{la}^{l}\} d^2\{\beta_{ha}^{l}\} \braket{\{\beta_{ka}^{l}\} | \{\beta_{ha}^{l}\}}  \bra{p_i, \{\beta_{ha}^{l}\}} S^{\dagger} \ket{p_l, \{\beta_{la}^{l}\}}  \braket{\{\beta_{la}^{l}\} | \{k_{j'}^{(\lambda_j')}\}_{n_{j'}}} \delta^{(3)}(p_{l} - p_{i'}) \ U \ket{p_i, \{\beta_{ka}^{l}\}} \Big]\\
\nn
&& = \frac{1}{\sqrt{D}} \Big(\frac{1}{\pi}\Big)^{4N} \sum_{\{n_{j'}\}} \int d^3p_{i'} d^3p_i \int d^2\{\beta_{ga}^{l}\} d^2\{\beta_{ka}^{l}\} d^2\{\beta_{la}^{l}\} d^2\{\beta_{ha}^{l}\} \braket{\{\beta_{ga}^{l}\} | \{k_{j'}^{(\lambda_j')}\}_{n_{j'}}} \braket{\{\beta_{ka}^{l}\} | \{\beta_{ha}^{l}\}} \braket{\{\beta_{la}^{l}\} | \{k_{j'}^{(\lambda_j')}\}_{n_{j'}}}\\
&& \bra{p_i, \{\beta_{ha}^{l}\}} S^{\dagger} \ket{p_{i'}, \{\beta_{la}^{l}\}} \ \ket{p_{i'}, \{\beta_{ga}^{l}\}} \otimes U \ket{p_i, \{\beta_{ka}^{l}\}}\,.
\ea
The density matrix associated casts
\ba
\nn
&& \rho_U = \ket{U} \bra{U}\\
\nn
&& = \frac{1}{D} \Big(\frac{1}{\pi}\Big)^{8N} \sum_{\{n_{j'}\}} \sum_{\{m_j\}} \int d^3p_{i'} d^3p_i d^3p_{r'} d^3p_r \int d^2\{\beta_{ga}^{l}\} d^2\{\beta_{ka}^{l}\} d^2\{\beta_{la}^{l}\} d^2\{\beta_{ha}^{l}\} d^2\{\beta_{g'a}^{l}\} d^2\{\beta_{k'a}^{l}\} d^2\{\beta_{l'a}^{l}\} d^2\{\beta_{h'a}^{l}\} \quad \ \\
\nn
&& \braket{\{\beta_{ga}^{l}\} | \{k_{j'}^{(\lambda_j')}\}_{n_{j'}}}_{in} \braket{\{\beta_{ka}^{l}\} | \{\beta_{ha}^{l}\}}  \bra{p_i, \{\beta_{ha}^{l}\}} S^{\dagger} \ket{p_{i'}, \{\beta_{la}^{l}\}} \braket{\{\beta_{la}^{l}\} | \{k_{j'}^{(\lambda_j')}\}_{n_{j'}}}_{in \ in} \braket{ \{k_{j}^{(\lambda_j)}\}_{m_j} | \{\beta_{g'a}^{l}\}} \braket{\{\beta_{h'a}^{l}\} | \{\beta_{k'a}^{l}\}}\\
&& \bra{p_{r'}, \{\beta_{l'a}^{l}\}} S \ket{p_r, \{\beta_{h'a}^{l}\}} \ _{in} \braket{\{k_{j}^{(\lambda_j)}\}_{m_j} | \{\beta_{l'a}^{l}\}} \ket{p_{i'}, \{\beta_{ga}^{l}\}} \bra{p_{r'}, \{\beta_{g'a}^{l}\}} \otimes \ U \ket{p_i, \{\beta_{ka}^{l}\}} \bra{p_r, \{\beta_{k'a}^{l}\}} U^{\dagger}\,.
\ea
In order to calculate the tripartite mutual information $I_{3(2)}$ in equation \eqref{10}, we have to construct the relevant reduced density matrix related to the states in \eqref{Choi-ce}.

The density matrix $\rho_U$ is essentially represented in terms of the states in $\mathcal{H}_{FQ} \otimes \mathcal{H}_{F\gamma}$. It can be further recast in terms of the states in $\mathcal{H}_{FQ} \otimes \mathcal{H}_{C\gamma}$. Finally, tracing out the relevant states in $\mathcal{H}_{FQ} \otimes \mathcal{H}_{F\gamma}$, rather than the states in $\mathcal{H}_{FQ} \otimes \mathcal{H}_{C\gamma}$, the following reduced density matrices, essential to our strategy, can be attained. Denoting with $A$ and $C$, respectively, the hard components of the incoming states, and outgoing states. And similarly denoting with $B$ and $D$, respectively, the soft components of the incoming states, and outgoing states. Then the states in $\mathcal{H}_{FQ} \otimes \mathcal{H}_{F\gamma}$, more precisely, the states in \eqref{ref19} and \eqref{ref21} can be label as\\
\ba
&& \ket{\alpha}_{in} = \ket{p_i, \{k_j^{(\lambda_j)}\}_{n_j}}_{in} = \ket{p_A,\{k_B\}},\\
&& \ket{\beta}_{out} = \ket{p_i, \{k_j^{(\lambda_j)}\}_{n_j}}_{out} = \ket{p_C,\{k_D\}}.
\ea
The reduced density matrix of $A$ and $C$ states is
\ba
\nn
&& \rho_{AC} = \tr_{BD} \rho_U\\
\nn
&& = \frac{1}{D} \Big(\frac{1}{\pi}\Big)^{8N} \sum_{\{n_{j'}\}} \sum_{\{m_j\}} \int d^3p_{i'} d^3p_i d^3p_{r'} d^3p_r \int d^2\{\beta_{ga}^{l}\} d^2\{\beta_{ka}^{l}\} d^2\{\beta_{la}^{l}\} d^2\{\beta_{ha}^{l}\} d^2\{\beta_{g'a}^{l}\} d^2\{\beta_{k'a}^{l}\} d^2\{\beta_{l'a}^{l}\} d^2\{\beta_{h'a}^{l}\} \quad \ \\
\nn
&& \braket{\{\beta_{ga}^{l}\} | \{k_{j'}^{(\lambda_j')}\}_{n_{j'}}}_{in} \braket{\{\beta_{ka}^{l}\} | \{\beta_{ha}^{l}\}}  \bra{p_i, \{\beta_{ha}^{l}\}} S^{\dagger} \ket{p_{i'}, \{\beta_{la}^{l}\}} \braket{\{\beta_{la}^{l}\} | \{k_{j'}^{(\lambda_j')}\}_{n_{j'}}}_{in \ in} \braket{ \{k_{j}^{(\lambda_j)}\}_{m_j} | \{\beta_{g'a}^{l}\}} \braket{\{\beta_{h'a}^{l}\} | \{\beta_{k'a}^{l}\}}\\
\nn
&& \bra{p_{r'}, \{\beta_{l'a}^{l}\}} S \ket{p_r, \{\beta_{h'a}^{l}\}} \ _{in} \braket{\{k_{j}^{(\lambda_j)}\}_{m_j} | \{\beta_{l'a}^{l}\}} \sum_{\{n'_s\}} \ _{in} \bra{\{k_s^{(\lambda_{s})}\}_{n'_s}} \Big[ \ket{p_{i'}, \{\beta_{ga}^{l}\}} \bra{p_{r'}, \{\beta_{g'a}^{l}\}} \Big] \ket{\{k_s^{(\lambda_{s})}\}_{n'_s}}_{in}\\
\nn
&& \otimes \sum_{\{m'_{s'}\}} \ _{out} \bra{\{k_{s'}^{(\lambda_{s'})}\}_{m'_{s'}}} \Big[ \sum_{\{m''_{s''}\}} \int d^3p_{f} \ket{p_{f}, \{k_{s''}^{(\lambda_{s''})}\}_{m''_{s''}}}_{out \ out} \bra{p_{f}, \{k_{s''}^{(\lambda_{s''})}\}_{m''_{s''}}} \Big] \Big[ U \ket{p_i, \{\beta_{ka}^{l}\}} \bra{p_r, \{\beta_{k'a}^{l}\}} U^{\dagger} \Big]\\
\nn
&& \Big[ \sum_{\{m'''_{s'''}\}} \int d^3p_{f'} \ket{p_{f'}, \{k_{s'''}^{(\lambda_{s'''})}\}_{m'''}}_{out \ out} \bra{p_{f'}, \{k_{s'''}^{(\lambda_{s'''})}\}_{m'''}} \Big] \ket{\{k_{s'}^{(\lambda_{s'})}\}_{m'_{s'}}}_{out}\\
\nn
&& = \frac{1}{D} \Big(\frac{1}{\pi}\Big)^{8N} \sum_{\{n_{j'}\}} \sum_{\{m_{j}\}} \int d^3p_{i'} d^3p_i d^3p_{r'} d^3p_r \int d^2\{\beta_{ga}^{l}\} d^2\{\beta_{ka}^{l}\} d^2\{\beta_{la}^{l}\} d^2\{\beta_{ha}^{l}\} d^2\{\beta_{g'a}^{l}\} d^2\{\beta_{k'a}^{l}\} d^2\{\beta_{l'a}^{l}\} d^2\{\beta_{h'a}^{l}\}\\
\nn
&& \braket{\{\beta_{ga}^{l}\} | \{k_{j'}^{(\lambda_j')}\}_{n_{j'}}}_{in} \braket{\{\beta_{ka}^{l}\} | \{\beta_{ha}^{l}\}}  \bra{p_i, \{\beta_{ha}^{l}\}} S^{\dagger} \ket{p_{i'}, \{\beta_{la}^{l}\}} \braket{\{\beta_{la}^{l}\} | \{k_{j'}^{(\lambda_j')}\}_{n_{j'}}}_{in \ in} \braket{ \{k_{j}^{(\lambda_j)}\}_{m_j} | \{\beta_{g'a}^{l}\}} \braket{\{\beta_{h'a}^{l}\} | \{\beta_{k'a}^{l}\}}\\
\nn
&& \bra{p_{r'}, \{\beta_{l'a}^{l}\}} S \ket{p_r, \{\beta_{h'a}^{l}\}} \ _{in} \braket{\{k_{j}^{(\lambda_j)}\}_{m_j} | \{\beta_{l'a}^{l}\}} \braket{\{\beta_{g'a}^{l}\} | \{\beta_{ga}^{l}\}} \braket{\{\beta_{k'a}^{l}\} | \{\beta_{ka}^{l}\}} \ \ket{p_{i'}}_{in \ in} \bra{p_{r'}} \otimes \ket{p_{i}}_{out \ out} \bra{p_{r}}\\
\nn
&& = \frac{1}{D} \Big(\frac{1}{\pi}\Big)^{4N} \sum_{\{n_{j'}\}} \sum_{\{m_{j}\}} \int d^3p_{i'} d^3p_i d^3p_{r'} d^3p_r \int d^2\{\beta_{i'a}^{l}\} d^2\{\beta_{ia}^{l}\} d^2\{\beta_{r'a}^{l}\} d^2\{\beta_{ra}^{l}\} \braket{\{\beta_{ra}^{l}\} | \{\beta_{ia}^{l}\}} \braket{p_i, \{\beta_{ia}^{l}\} | S^{\dagger} | p_{i'}, \{\beta_{i'a}^{l}\}}\\
\nn
&& \braket{p_{r'}, \{\beta_{r'a}^{l}\} | S | p_r, \{\beta_{ra}^{l}\}} \braket{\{\beta_{i'a}^{l}\} | \{k_{j'}^{(\lambda_j')}\}_{n_{j'}}}_{in \ in} \braket{ \{k_{j}^{(\lambda_j)}\}_{m_j} | \{k_{j'}^{(\lambda_j')}\}_{n_{j'}}}_{in \ in} \braket{\{k_{j}^{(\lambda_j)}\}_{m_j} | \{\beta_{r'a}^{l}\}} \ \ket{p_{i'}}_{in \ in} \bra{p_{r'}}\\
\nn
&& \otimes \ket{p_{i}}_{out \ out} \bra{p_{r}}\\
\nn
&& = \frac{1}{D} \Big(\frac{1}{\pi}\Big)^{4N} \int d^3p_{i'} d^3p_i d^3p_{r'} d^3p_r \int d^2\{\beta_{i'a}^{l}\} d^2\{\beta_{ia}^{l}\} d^2\{\beta_{r'a}^{l}\} d^2\{\beta_{ra}^{l}\} \braket{p_i, \{\beta_{ia}^{l}\} | S^{\dagger} | p_{i'}, \{\beta_{i'a}^{l}\}} \braket{\{\beta_{i'a}^{l}\} | \{\beta_{r'a}^{l}\}}\\
\nn
&& \braket{p_{r'}, \{\beta_{r'a}^{l}\} | S | p_r, \{\beta_{ra}^{l}\}} \braket{\{\beta_{ra}^{l}\} | \{\beta_{ia}^{l}\}} \ \ket{p_{i'}}_{in \ in} \bra{p_{r'}} \otimes \ \ket{p_{i}}_{out \ out} \bra{p_{r}}\\
\nn
&& = \frac{1}{D} \Big(\frac{1}{\pi}\Big)^{2N} \int d^3p_{i'} d^3p_i d^3p_{r'} d^3p_{r} \int d^2\{\beta_{i'a}^{l}\} d^2\{\beta_{ia}^{l}\} \braket{p_i, \{\beta_{ia}^{l}\} | S^{\dagger} | p_{i'}, \{\beta_{i'a}^{l}\}} \braket{p_{r'}, \{\beta_{i'a}^{l}\} | S | p_r, \{\beta_{ia}^{l}\}} \ \ket{p_{i'}}_{in \ in} \bra{p_{r'}}\\
&& \otimes \ \ket{p_{i}}_{out \ out} \bra{p_{r}}\,,
\ea
where states $B$, which denote the soft component of the incoming states, and states $D$, the outgoing states, have been traced out.\\

The reduced density matrix of the $C$ states is  recovered by tracing out the hard part components $A$ of the incoming states, 
\ba
\nn
&& \rho_{C} = \tr_{A} \rho_{AC}\\
\nn
&& = \frac{1}{D} \Big(\frac{1}{\pi}\Big)^{2N} \int d^3p_{i'} d^3p_i d^3p_{r} \int d^2\{\beta_{i'a}^{l}\} d^2\{\beta_{ia}^{l}\} \braket{p_i, \{\beta_{ia}^{l}\} | S^{\dagger} | p_{i'}, \{\beta_{i'a}^{l}\}} \braket{p_{i'}, \{\beta_{i'a}^{l}\} | S | p_r, \{\beta_{ia}^{l}\}} \ \ket{p_{i}}_{out \ out} \bra{p_{r}} \quad \quad \quad\\
&& = \frac{1}{D} \Big(\frac{1}{\pi}\Big)^{N} \int d^3p_i d^3p_{r} \int d^2\{\beta_{ia}^{l}\} \braket{\{\beta_{ia}^{l}\} | \{\beta_{ia}^{l}\}} \delta^{(3)}(p_{i} - p_{r}) \ \ket{p_{i}}_{out \ out} \bra{p_{r}} = \frac{d_{B}}{D} \int d^3p_i \ \ket{p_{i}}_{out \ out} \bra{p_{i}}.
\ea

Similarly, tracing $B$ and $C$ states, respectively ongoing soft photons and outgoing hard part charged particle, we obtain for the density matrix related to incoming hard charged particles $A$ and outgoing soft photons $D$
\ba
\nn
&& \rho_{AD} = \tr_{BC} \rho_U\\
\nn
&& = \frac{1}{D} \Big(\frac{1}{\pi}\Big)^{8N} \sum_{\{n_{j'}\}} \sum_{\{m_{j}\}} \int d^3p_{i'} d^3p_i d^3p_{r'} d^3p_r \int d^2\{\beta_{ga}^{l}\} d^2\{\beta_{ka}^{l}\} d^2\{\beta_{la}^{l}\} d^2\{\beta_{ha}^{l}\} d^2\{\beta_{g'a}^{l}\} d^2\{\beta_{k'a}^{l}\} d^2\{\beta_{l'a}^{l}\} d^2\{\beta_{h'a}^{l}\}\\
\nn
&& \braket{\{\beta_{ga}^{l}\} | \{k_{j'}^{(\lambda_j')}\}_{n_{j'}}}_{in} \braket{\{\beta_{ka}^{l}\} | \{\beta_{ha}^{l}\}}  \bra{p_i, \{\beta_{ha}^{l}\}} S^{\dagger} \ket{p_{i'}, \{\beta_{la}^{l}\}} \braket{\{\beta_{la}^{l}\} | \{k_{j'}^{(\lambda_j')}\}_{n_{j'}}}_{in \ in} \braket{ \{k_{j}^{(\lambda_j)}\}_{m_j} | \{\beta_{g'a}^{l}\}} \braket{\{\beta_{h'a}^{l}\} | \{\beta_{k'a}^{l}\}}\\
\nn
&& \bra{p_{r'}, \{\beta_{l'a}^{l}\}} S \ket{p_r, \{\beta_{h'a}^{l}\}} \ _{in} \braket{\{k_{j}^{(\lambda_j)}\}_{m_j} | \{\beta_{l'a}^{l}\}} \sum_{\{n'_s\}} \ _{in} \bra{\{k_s^{(\lambda_{s})}\}_{n'_s}} \Big[ \ket{p_{i'}, \{\beta_{ga}^{l}\}} \bra{p_{r'}, \{\beta_{g'a}^{l}\}} \Big] \ket{\{k_s^{(\lambda_{s})}\}_{n'_s}}_{in}\\
\nn
&& \otimes \int d^3p_{f} \ _{out}\bra{p_{f}} \Big[ \sum_{\{m'_{s'}\}} \int d^3p_{f'} \ket{p_{f'}, \{k_{s'}^{(\lambda_{s'})}\}_{m'_{s'}}}_{out \ out} \bra{p_{f'}, \{k_{s'}^{(\lambda_{s'})}\}_{m'_{s'}}} \Big] \Big[ U \ket{p_i, \{\beta_{ka}^{l}\}} \bra{p_r, \{\beta_{k'a}^{l}\}} U^{\dagger} \Big]\\
\nn
&& \Big[ \sum_{\{m''_{s''}\}} \int d^3p_{f''} \ket{p_{f''}, \{k_{s''}^{(\lambda_{s''})}\}_{m''_{s''}}}_{out \ out} \bra{p_{f''}, \{k_{s''}^{(\lambda_{s''})}\}_{m''_{s''}}} \Big] \ket{p_{f}}_{out}\\
\nn
&& = \frac{1}{D} \Big(\frac{1}{\pi}\Big)^{4N} \sum_{\{n_{j'}\}} \sum_{\{m_{j}\}} \sum_{\{m'_{s'}\}} \sum_{\{m''_{s''}\}} \int d^3p_{i'} d^3p_i d^3p_{r'} \int d^2\{\beta_{i'a}^{l}\} d^2\{\beta_{ia}^{l}\} d^2\{\beta_{r'a}^{l}\} d^2\{\beta_{h'a}^{l}\} \bra{p_i, \{\beta_{ia}^{l}\}} S^{\dagger} \ket{p_{i'}, \{\beta_{i'a}^{l}\}}\\
\nn
&& \bra{p_{r'}, \{\beta_{r'a}^{l}\}} S \ket{p_i, \{\beta_{h'a}^{l}\}} \braket{\{\beta_{i'a}^{l}\} | \{k_{j'}^{(\lambda_j')}\}_{n_{j'}}}_{in \ in} \braket{ \{k_{j}^{(\lambda_j)}\}_{m_j} | \{k_{j'}^{(\lambda_j')}\}_{n_{j'}}}_{in \ in} \braket{\{k_{j}^{(\lambda_j)}\}_{m_j} | \{\beta_{r'a}^{l}\}} \braket{\{\beta_{h'a}^{l}\} | \{k_{s''}^{(\lambda_{s''})}\}_{m''_{s''}}}_{in}\\
\nn
&& \ _{in} \braket{\{k_{s'}^{(\lambda_{s'})}\}_{m'_{s'}} | \{\beta_{ia}^{l}\}} \ \ket{p_{i'}}_{in \ in} \bra{p_{r'}} \otimes \ \ket{\{k_{s'}^{(\lambda_{s'})}\}_{m'_{s'}}}_{out \ out} \bra{\{k_{s''}^{(\lambda_{s''})}\}_{m''_{s''}}}\\
\nn
&& = \frac{1}{D} \Big(\frac{1}{\pi}\Big)^{4N} \sum_{\{m'_{s'}\}} \sum_{\{m''_{s''}\}} \int d^3p_{i'} d^3p_i d^3p_{r'} \int d^2\{\beta_{i'a}^{l}\} d^2\{\beta_{ia}^{l}\} d^2\{\beta_{r'a}^{l}\} d^2\{\beta_{h'a}^{l}\} \braket{p_i, \{\beta_{ia}^{l}\} | S^{\dagger} | p_{i'}, \{\beta_{i'a}^{l}\}} \braket{\{\beta_{i'a}^{l}\} | \{\beta_{r'a}^{l}\}}\\
\nn
&& \braket{p_{r'}, \{\beta_{r'a}^{l}\} | S | p_i, \{\beta_{h'a}^{l}\}} \braket{\{\beta_{h'a}^{l}\} | \{k_{s''}^{(\lambda_{s''})}\}_{m''_{s''}}}_{in \ in} \braket{\{k_{s'}^{(\lambda_{s'})}\}_{m'_{s'}} | \{\beta_{ia}^{l}\}} \ \ket{p_{i'}}_{in \ in} \bra{p_{r'}} \otimes \ket{\{k_{s'}^{(\lambda_{s'})}\}_{m'_{s'}}}_{out \ out} \bra{\{k_{s''}^{(\lambda_{s''})}\}_{m''_{s''}}}\\
\nn
&& = \frac{1}{D} \Big(\frac{1}{\pi}\Big)^{3N} \sum_{\{m'_{s'}\}} \sum_{\{m''_{s''}\}} \int d^3p_{i'} d^3p_i d^3p_{r'} \int d^2\{\beta_{i'a}^{l}\} d^2\{\beta_{ia}^{l}\} d^2\{\beta_{h'a}^{l}\} \braket{p_i, \{\beta_{ia}^{l}\} | S^{\dagger} | p_{i'}, \{\beta_{i'a}^{l}\}} \braket{p_{r'}, \{\beta_{i'a}^{l}\} | S | p_i, \{\beta_{h'a}^{l}\}}\\
&& \braket{\{\beta_{h'a}^{l}\} | \{k_{s''}^{(\lambda_{s''})}\}_{m''}}_{in \ in} \braket{\{k_{s'}^{(\lambda_{s'})}\}_{m'} | \{\beta_{ia}^{l}\}} \ \ket{p_{i'}}_{in \ in} \bra{p_{r'}} \otimes \ket{\{k_{s'}^{(\lambda_{s'})}\}_{m'}}_{out \ out} \bra{\{k_{s''}^{(\lambda_{s''})}\}_{m''}},
\ea
finally, tracing out incoming charged particle $A$, we derive for outgoing soft particles $D$
\ba
\nn
&& \rho_{D} = \tr_{A} \rho_{AD}\\
\nn
&& = \frac{1}{D} \Big(\frac{1}{\pi}\Big)^{3N} \sum_{\{m'_{s'}\}} \sum_{\{m''_{s''}\}} \int d^3p_{i'} d^3p_i \int d^2\{\beta_{i'a}^{l}\} d^2\{\beta_{ia}^{l}\} d^2\{\beta_{h'a}^{l}\} \braket{p_i, \{\beta_{ia}^{l}\} | S^{\dagger} | p_{i'}, \{\beta_{i'a}^{l}\}} \braket{p_{i'}, \{\beta_{i'a}^{l}\} | S | p_i, \{\beta_{h'a}^{l}\}} \qquad\\
\nn
&& \braket{\{\beta_{h'a}^{l}\} | \{k_{s''}^{(\lambda_{s''})}\}_{m''_{s''}}}_{in \ in} \braket{\{k_{s'}^{(\lambda_{s'})}\}_{m'_{s'}} | \{\beta_{ia}^{l}\}} \ \ket{\{k_{s'}^{(\lambda_{s'})}\}_{m'_{s'}}}_{out \ out} \bra{\{k_{s''}^{(\lambda_{s''})}\}_{m''_{s''}}}\\
\nn
&& = \frac{1}{D} \Big(\frac{1}{\pi}\Big)^{2N} \sum_{\{m'_{s'}\}} \sum_{\{m''_{s''}\}} \int d^3p_i \int d^2\{\beta_{ia}^{l}\} d^2\{\beta_{h'a}^{l}\} \delta^{(3)}(0) \braket{\{\beta_{ia}^{l}\} | \{\beta_{h'a}^{l}\}} \braket{\{\beta_{h'a}^{l}\} | \{k_{s''}^{(\lambda_{s''})}\}_{m''_{s''}}}_{in \ in} \braket{\{k_{s'}^{(\lambda_{s'})}\}_{m'_{s'}} | \{\beta_{ia}^{l}\}}\\
\nn
&& \ket{\{k_{s'}^{(\lambda_{s'})}\}_{m'_{s'}}}_{out \ out} \bra{\{k_{s''}^{(\lambda_{s''})}\}_{m''_{s''}}}\\
\nn
&& = \frac{1}{D} \sum_{\{m'_{s'}\}} \sum_{\{m''_{s''}\}} \int d^3p_i \delta^{(3)}(0) \ _{in} \braket{\{k_{s'}^{(\lambda_{s'})}\}_{m'_{s'}} | \{k_{s''}^{(\lambda_{s''})}\}_{m''_{s''}}}_{in} \ \ket{\{k_{s'}^{(\lambda_{s'})}\}_{m'_{s'}}}_{out \ out} \bra{\{k_{s''}^{(\lambda_{s''})}\}_{m''_{s''}}}\\
&& = \frac{d_{A}}{D} \sum_{\{m'_{s'}\}} \ket{\{k_{s'}^{(\lambda_{s'})}\}_{m'_{s'}}}_{out \ out} \bra{\{k_{s'}^{(\lambda_{s'})}\}_{m'_{s'}}}\,.
\ea

It is now straightforward to compute the 2-Renyi entropies related to the aforementioned quantities:
\ba
\nn
&& S_2(AC) = -\log\tr\rho_{AC}^2\\
\nn
&& = -\log\tr \Big[ \frac{1}{D^2} \Big(\frac{1}{\pi}\Big)^{4N} \int d^3p_{i'} d^3p_{i} d^3p_{r'} d^3p_{r} d^3p_{h'} d^3p_{h} \int d^2\{\beta_{i'a}^{l}\} d^2\{\beta_{ia}^{l}\} d^2\{\beta_{g'a}^{l}\} d^2\{\beta_{ga}^{l}\} \braket{p_i, \{\beta_{ia}^{l}\} | S^{\dagger} | p_{i'}, \{\beta_{i'a}^{l}\}}\\
\nn
&& \braket{p_{r'}, \{\beta_{i'a}^{l}\} | S | p_r, \{\beta_{ia}^{l}\}} \braket{p_r, \{\beta_{ga}^{l}\} | S^{\dagger} | p_{r'}, \{\beta_{g'a}^{l}\}} \braket{p_{h'}, \{\beta_{g'a}^{l}\} | S | p_h, \{\beta_{ga}^{l}\}} \ \ket{p_{i'}}_{in \ in} \bra{p_{h'}} \otimes \ket{p_{i}}_{out \ out} \bra{p_{h}} \Big]\\
\nn
&& = -\log \Big[ \frac{1}{D^2} \Big(\frac{1}{\pi}\Big)^{4N} \int d^3p_{i'} d^3p_{i} d^3p_{r'} d^3p_{r} \int d^2\{\beta_{i'a}^{l}\} d^2\{\beta_{ia}^{l}\} d^2\{\beta_{r'a}^{l}\} d^2\{\beta_{ra}^{l}\} \braket{p_i, \{\beta_{ia}^{l}\} | S^{\dagger} | p_{i'}, \{\beta_{i'a}^{l}\}}\\
\nn
&& \braket{p_{r'}, \{\beta_{i'a}^{l}\} | S | p_r, \{\beta_{ia}^{l}\}} \braket{p_r, \{\beta_{ra}^{l}\} | S^{\dagger} | p_{r'}, \{\beta_{r'a}^{l}\}} \braket{p_{i'}, \{\beta_{r'a}^{l}\} | S | p_i, \{\beta_{ra}^{l}\}} \Big]\\
\nn
&& = -\log \Big[ \frac{1}{D^2} \Big(\frac{1}{\pi}\Big)^{8N} \int d^3p_{i'} d^3p_i d^3p_{r'} d^3p_r \int d^2\{\beta_{i'a}^{l}\} d^2\{\beta_{ia}^{l}\} d^2\{\beta_{r'a}^{l}\} d^2\{\beta_{ra}^{l}\} d^2\{\beta_{g'a}^{l}\} d^2\{\beta_{ga}^{l}\} d^2\{\beta_{h'a}^{l}\} d^2\{\beta_{ha}^{l}\} \quad \\
\nn
&& \braket{p_i, \{\beta_{ia}^{l}\} | S^{\dagger} | p_{i'}, \{\beta_{i'a}^{l}\}} \braket{\{\beta_{i'a}^{l}\} | \{\beta_{r'a}^{l}\}} \braket{p_{r'}, \{\beta_{r'a}^{l}\} | S | p_r, \{\beta_{ra}^{l}\}} \braket{\{\beta_{ra}^{l}\} | \{\beta_{ia}^{l}\}} \braket{p_r, \{\beta_{ga}^{l}\} | S^{\dagger} | p_{r'}, \{\beta_{g'a}^{l}\}}\\
&& \braket{\{\beta_{g'a}^{l}\} | \{\beta_{h'a}^{l}\}} \braket{p_{i'}, \{\beta_{h'a}^{l}\} | S | p_i, \{\beta_{ha}^{l}\}} \braket{\{\beta_{ha}^{l}\} | \{\beta_{ga}^{l}\}} \Big];
\ea

and 

\be
S_2(C) = -\log\tr\rho_{C}^2 = -\log\tr \Big[ \frac{d_{B}^2}{D^2} \int d^3p_i \ \ket{p_{i}}_{out \ out} \bra{p_{i}} \Big] = -\log \Big[ \frac{d_{B}^2d_{A}}{D^2} \Big] = \log d_{A}; \qquad \qquad \qquad \qquad \quad \quad
\ee

and 

\ba
\nn
&& S_2(AD) = -\log\tr\rho_{AD}^2\\
\nn
&& = -\log\tr \Big[ \frac{1}{D^2} \Big(\frac{1}{\pi}\Big)^{6N} \sum_{\{m'_{s'}\}} \sum_{\{m''_{s''}\}} \sum_{\{n''_{j''}\}} \int d^3p_{i'} d^3p_i d^3p_{r'} d^3p_g d^3p_{r} \int d^2\{\beta_{i'a}^{l}\} d^2\{\beta_{ia}^{l}\} d^2\{\beta_{h'a}^{l}\} d^2\{\beta_{g'a}^{l}\} d^2\{\beta_{ga}^{l}\} \ \ \ \ \\
\nn
&& d^2\{\beta_{ha}^{l}\} \braket{p_i, \{\beta_{ia}^{l}\} | S^{\dagger} | p_{i'}, \{\beta_{i'a}^{l}\}} \braket{p_{r'}, \{\beta_{i'a}^{l}\} | S | p_i, \{\beta_{h'a}^{l}\}} \braket{p_g, \{\beta_{ga}^{l}\} | S^{\dagger} | p_{r'}, \{\beta_{g'a}^{l}\}} \braket{p_{r}, \{\beta_{g'a}^{l}\} | S | p_g, \{\beta_{ha}^{l}\}}\\
\nn
&& \braket{\{\beta_{h'a}^{l}\} | \{k_{s''}^{(\lambda_{s''})}\}_{m''_{s''}}}_{in \ in} \braket{\{k_{s''}^{(\lambda_{s''})}\}_{m''_{s''}} | \{\beta_{ga}^{l}\}} \braket{\{\beta_{ha}^{l}\} | \{k_{j''}^{(\lambda_{j''})}\}_{n''_{j''}}}_{in \ in} \braket{\{k_{s'}^{(\lambda_{s'})}\}_{m'_{s'}} | \{\beta_{ia}^{l}\}} \ \ket{p_{i'}}_{in \ in} \bra{p_{r}}\\
\nn
&& \otimes \ket{\{k_{s'}^{(\lambda_{s'})}\}_{m'_{s'}}}_{out \ out} \bra{\{k_{j''}^{(\lambda_{j''})}\}_{n''_{j''}}} \Big]\\
\nn
&& = -\log\tr \Big[ \frac{1}{D^2} \Big(\frac{1}{\pi}\Big)^{6N} \sum_{\{m'_{s'}\}} \sum_{\{n''_{j''}\}} \int d^3p_{i'} d^3p_i d^3p_{r'} d^3p_g d^3p_{r} \int d^2\{\beta_{i'a}^{l}\} d^2\{\beta_{ia}^{l}\} d^2\{\beta_{h'a}^{l}\} d^2\{\beta_{g'a}^{l}\} d^2\{\beta_{ga}^{l}\} d^2\{\beta_{ha}^{l}\}\\
\nn
&& \braket{p_i, \{\beta_{ia}^{l}\} | S^{\dagger} | p_{i'}, \{\beta_{i'a}^{l}\}} \braket{p_{r'}, \{\beta_{i'a}^{l}\} | S | p_i, \{\beta_{h'a}^{l}\}} \braket{p_g, \{\beta_{ga}^{l}\} | S^{\dagger} | p_{r'}, \{\beta_{g'a}^{l}\}} \braket{p_{r}, \{\beta_{g'a}^{l}\} | S | p_g, \{\beta_{ha}^{l}\}} \braket{\{\beta_{h'a}^{l}\} | \{\beta_{ga}^{l}\}}\\
\nn
&& \braket{\{\beta_{ha}^{l}\} | \{k_{j''}^{(\lambda_{j''})}\}_{n''_{j''}}}_{in \ in} \braket{\{k_{s'}^{(\lambda_{s'})}\}_{m'_{s'}} | \{\beta_{ia}^{l}\}} \ \ket{p_{i'}}_{in \ in} \bra{p_{r}} \otimes \ket{\{k_{s'}^{(\lambda_{s'})}\}_{m'_{s'}}}_{out \ out} \bra{\{k_{j''}^{(\lambda_{j''})}\}_{n''_{j''}}} \Big]\\
\nn
&& = -\log \Big[ \frac{1}{D^2} \Big(\frac{1}{\pi}\Big)^{6N} \sum_{\{m'_{s'}\}} \int d^3p_{i'} d^3p_i d^3p_{r'} d^3p_g \int d^2\{\beta_{i'a}^{l}\} d^2\{\beta_{ia}^{l}\} d^2\{\beta_{h'a}^{l}\} d^2\{\beta_{g'a}^{l}\} d^2\{\beta_{ga}^{l}\} d^2\{\beta_{ha}^{l}\}\\
\nn
&& \braket{p_i, \{\beta_{ia}^{l}\} | S^{\dagger} | p_{i'}, \{\beta_{i'a}^{l}\}} \braket{p_{r'}, \{\beta_{i'a}^{l}\} | S | p_i, \{\beta_{h'a}^{l}\}} \braket{p_g, \{\beta_{ga}^{l}\} | S^{\dagger} | p_{r'}, \{\beta_{g'a}^{l}\}} \braket{p_{i'}, \{\beta_{g'a}^{l}\} | S | p_g, \{\beta_{ha}^{l}\}} \braket{\{\beta_{h'a}^{l}\} | \{\beta_{ga}^{l}\}}\\
\nn
&& \braket{\{\beta_{ha}^{l}\} | \{k_{s'}^{(\lambda_{s'})}\}_{m'_{s'}}}_{in \ in} \braket{\{k_{s'}^{(\lambda_{s'})}\}_{m'_{s'}} | \{\beta_{ia}^{l}\}} \Big]\\
\nn
&& = -\log \Big[ \frac{1}{D^2} \Big(\frac{1}{\pi}\Big)^{4N} \int d^3p_{i'} d^3p_i d^3p_{r'} d^3p_r \int d^2\{\beta_{i'a}^{l}\} d^2\{\beta_{ia}^{l}\} d^2\{\beta_{r'a}^{l}\} d^2\{\beta_{ra}^{l}\} \braket{p_i, \{\beta_{ia}^{l}\} | S^{\dagger} | p_{i'}, \{\beta_{i'a}^{l}\}}\\
\nn
&& \braket{p_{r'}, \{\beta_{i'a}^{l}\} | S | p_i, \{\beta_{ra}^{l}\}} \braket{p_r, \{\beta_{ra}^{l}\} | S^{\dagger} | p_{r'}, \{\beta_{r'a}^{l}\}} \braket{p_{i'}, \{\beta_{r'a}^{l}\} | S | p_r, \{\beta_{ia}^{l}\}} \Big]\\
\nn
&& = - \log \Big[ \frac{1}{D^2} \Big(\frac{1}{\pi}\Big)^{8N} \int d^3p_{i'} d^3p_i d^3p_{r'} d^3p_g \int d^2\{\beta_{i'a}^{l}\} d^2\{\beta_{ia}^{l}\} d^2\{\beta_{r'a}^{l}\} d^2\{\beta_{fa}^{l}\} d^2\{\beta_{g'a}^{l}\} d^2\{\beta_{ga}^{l}\} d^2\{\beta_{h'a}^{l}\} d^2\{\beta_{la}^{l}\} \quad\\
\nn
&& \braket{p_i, \{\beta_{ia}^{l}\} | S^{\dagger} | p_{i'}, \{\beta_{i'a}^{l}\}} \braket{\{\beta_{i'a}^{l}\} | \{\beta_{r'a}^{l}\}} \braket{p_{r'}, \{\beta_{r'a}^{l}\} | S | p_i, \{\beta_{fa}^{l}\}} \braket{\{\beta_{fa}^{l}\} | \{\beta_{ga}^{l}\}} \braket{p_g, \{\beta_{ga}^{l}\} | S^{\dagger} | p_{r'}, \{\beta_{g'a}^{l}\}}\\
&& \braket{\{\beta_{g'a}^{l}\} | \{\beta_{h'a}^{l}\}} \braket{p_{i'}, \{\beta_{h'a}^{l}\} | S | p_g, \{\beta_{la}^{l}\}} \braket{\{\beta_{la}^{l}\} | \{\beta_{ia}^{l}\}} \Big]\,;
\ea

and 

\be
S_2(D) = -\log\tr\rho_{D}^2 = -\log\tr \Big[ \frac{d_{A}^2}{D^2} \sum_{\{m'_{s'}\}} \ket{\{k_{s'}^{(\lambda_{s'})}\}_{m'_{s'}}}_{out \ out} \bra{\{k_{s'}^{(\lambda_{s'})}\}_{m'_{s'}}} \Big] = -\log \Big[ \frac{d_{A}^2d_{B}}{D^2} \Big] = \log d_{B}. \quad \quad \quad
\ee

Once these contributions have been evaluated, the tripartite mutual information $I_{3(2)}$ can be expressed as
\ba
\nn
&& I_{3(2)} = S_2(C) + S_2(D) - S_2(AC) - S_2(AD)\\
\nn
&& = \log d_{A} + \log d_{B} + \log \Big[ \frac{1}{D^2} \Big(\frac{1}{\pi}\Big)^{8N} \int d^3p_{i'} d^3p_i d^3p_{r'} d^3p_r \int d^2\{\beta_{i'a}^{l}\} d^2\{\beta_{ia}^{l}\} d^2\{\beta_{r'a}^{l}\} d^2\{\beta_{ra}^{l}\} d^2\{\beta_{g'a}^{l}\}\\
\nn
&& d^2\{\beta_{ga}^{l}\} d^2\{\beta_{h'a}^{l}\} d^2\{\beta_{ha}^{l}\} \braket{p_i, \{\beta_{ia}^{l}\} | S^{\dagger} | p_{i'}, \{\beta_{i'a}^{l}\}} \braket{\{\beta_{i'a}^{l}\} | \{\beta_{r'a}^{l}\}} \braket{p_{r'}, \{\beta_{r'a}^{l}\} | S | p_r, \{\beta_{ra}^{l}\}} \braket{\{\beta_{ra}^{l}\} | \{\beta_{ia}^{l}\}}\\
\nn
&& \braket{p_r, \{\beta_{ga}^{l}\} | S^{\dagger} | p_{r'}, \{\beta_{g'a}^{l}\}} \braket{\{\beta_{g'a}^{l}\} | \{\beta_{h'a}^{l}\}} \braket{p_{i'}, \{\beta_{h'a}^{l}\} | S | p_i, \{\beta_{ha}^{l}\}} \braket{\{\beta_{ha}^{l}\} | \{\beta_{ga}^{l}\}} \Big] + \log \Big[ \frac{1}{D^2} \Big(\frac{1}{\pi}\Big)^{8N}\\
\nn
&& \int d^3p_{i'} d^3p_i d^3p_{r'} d^3p_g \int d^2\{\beta_{i'a}^{l}\} d^2\{\beta_{ia}^{l}\} d^2\{\beta_{r'a}^{l}\} d^2\{\beta_{fa}^{l}\} d^2\{\beta_{g'a}^{l}\} d^2\{\beta_{ga}^{l}\} d^2\{\beta_{h'a}^{l}\} d^2\{\beta_{la}^{l}\} \braket{\{\beta_{la}^{l}\} | \{\beta_{ia}^{l}\}}\\
\nn
&& \braket{p_i, \{\beta_{ia}^{l}\} | S^{\dagger} | p_{i'}, \{\beta_{i'a}^{l}\}} \braket{\{\beta_{i'a}^{l}\} | \{\beta_{r'a}^{l}\}} \braket{p_{r'}, \{\beta_{r'a}^{l}\} | S | p_i, \{\beta_{fa}^{l}\}} \braket{\{\beta_{fa}^{l}\} | \{\beta_{ga}^{l}\}} \braket{p_g, \{\beta_{ga}^{l}\} | S^{\dagger} | p_{r'}, \{\beta_{g'a}^{l}\}}\\
&& \braket{\{\beta_{g'a}^{l}\} | \{\beta_{h'a}^{l}\}} \braket{p_{i'}, \{\beta_{h'a}^{l}\} | S | p_g, \{\beta_{la}^{l}\}} \Big].
\ea

This a rather convoluted expression, which nevertheless will allow us to draw relevant  consequences in the next sections.

\subsection{Tripartite mutual information at the leading-order}
\noindent
The expression of the $S$ matrix generic element $\braket{p_{i'}, \{\beta_{i'a}^{l}\} | S | p_{i}, \{\beta_{ia}^{l}\}}$ in \eqref{46} has been expressed previously. We deploy now that result, substituting it into the integrand of $I_{3(2)}$, and finally expand it at the leading-order. We then obtain:

\ba
\nn
&& \braket{p_i, \{\beta_{ia}^{l}\} | S^{\dagger} | p_{i'}, \{\beta_{i'a}^{l}\}} \braket{\{\beta_{i'a}^{l}\} | \{\beta_{r'a}^{l}\}} \braket{p_{r'}, \{\beta_{r'a}^{l}\} | S | p_r, \{\beta_{ra}^{l}\}} \braket{\{\beta_{ra}^{l}\} | \{\beta_{ia}^{l}\}} \braket{p_r, \{\beta_{ga}^{l}\} | S^{\dagger} | p_{r'}, \{\beta_{g'a}^{l}\}}\\
\nn
&& \braket{\{\beta_{g'a}^{l}\} | \{\beta_{h'a}^{l}\}} \braket{p_{i'}, \{\beta_{h'a}^{l}\} | S | p_i, \{\beta_{ha}^{l}\}} \braket{\{\beta_{ha}^{l}\} | \{\beta_{ga}^{l}\}}\\
\nn
&& \simeq \delta^{(3)}(p_{i} - p_{i'}) \delta^{(3)}(p_{r'} - p_{r}) \delta^{(3)}(p_{r} - p_{r'}) \delta^{(3)}(p_{i'} - p_{i}) \braket{\{\beta_{ia}^{l}\} | \{\beta_{i'a}^{l}\}} \braket{\{\beta_{r'a}^{l}\} | \{\beta_{ra}^{l}\}} \braket{\{\beta_{ga}^{l}\} | \{\beta_{g'a}^{l}\}} \braket{\{\beta_{h'a}^{l}\} | \{\beta_{ha}^{l}\}}\\
\nn
&& \braket{\{\beta_{i'a}^{l}\} | \{\beta_{r'a}^{l}\}} \braket{\{\beta_{ra}^{l}\} | \{\beta_{ia}^{l}\}} \braket{\{\beta_{g'a}^{l}\} | \{\beta_{h'a}^{l}\}} \braket{\{\beta_{ha}^{l}\} | \{\beta_{ga}^{l}\}}\\
\nn
&& + \ 2i\pi \delta(p_{i}^0 - p_{i'}^0) \braket{p_{i}, \{\beta_{ia}^{l}\} | M_e^{\dagger} | p_{i'}, \{\beta_{i'a}^{l}\}} \delta^{(3)}(p_{r'} - p_{r}) \delta^{(3)}(p_{r} - p_{r'}) \delta^{(3)}(p_{i'} - p_{i}) \braket{\{\beta_{r'a}^{l}\} | \{\beta_{ra}^{l}\}} \braket{\{\beta_{ga}^{l}\} | \{\beta_{g'a}^{l}\}}\\
\nn
&& \braket{\{\beta_{h'a}^{l}\} | \{\beta_{ha}^{l}\}} \braket{\{\beta_{i'a}^{l}\} | \{\beta_{r'a}^{l}\}} \braket{\{\beta_{ra}^{l}\} | \{\beta_{ia}^{l}\}} \braket{\{\beta_{g'a}^{l}\} | \{\beta_{h'a}^{l}\}} \braket{\{\beta_{ha}^{l}\} | \{\beta_{ga}^{l}\}}\\
\nn
&& - \ 2i\pi \delta(p_{r'}^0 - p_{r}^0) \braket{p_{r'}, \{\beta_{r'a}^{l}\} | M_e | p_{r}, \{\beta_{ra}^{l}\}} \delta^{(3)}(p_{i} - p_{i'}) \delta^{(3)}(p_{r} - p_{r'}) \delta^{(3)}(p_{i'} - p_{i}) \braket{\{\beta_{ia}^{l}\} | \{\beta_{i'a}^{l}\}} \braket{\{\beta_{ga}^{l}\} | \{\beta_{g'a}^{l}\}}\\
\nn
&& \braket{\{\beta_{h'a}^{l}\} | \{\beta_{ha}^{l}\}} \braket{\{\beta_{i'a}^{l}\} | \{\beta_{r'a}^{l}\}} \braket{\{\beta_{ra}^{l}\} | \{\beta_{ia}^{l}\}} \braket{\{\beta_{g'a}^{l}\} | \{\beta_{h'a}^{l}\}} \braket{\{\beta_{ha}^{l}\} | \{\beta_{ga}^{l}\}}\\
\nn
&& + \ 2i\pi \delta(p_{r}^0 - p_{r'}^0) \braket{p_r, \{\beta_{ga}^{l}\} | M_e^{\dagger} | p_{r'}, \{\beta_{g'a}^{l}\}} \delta^{(3)}(p_{i} - p_{i'}) \delta^{(3)}(p_{r'} - p_{r}) \delta^{(3)}(p_{i'} - p_{i}) \braket{\{\beta_{ia}^{l}\} | \{\beta_{i'a}^{l}\}} \braket{\{\beta_{r'a}^{l}\} | \{\beta_{ra}^{l}\}}\\
\nn
&& \braket{\{\beta_{h'a}^{l}\} | \{\beta_{ha}^{l}\}} \braket{\{\beta_{i'a}^{l}\} | \{\beta_{r'a}^{l}\}} \braket{\{\beta_{ra}^{l}\} | \{\beta_{ia}^{l}\}} \braket{\{\beta_{g'a}^{l}\} | \{\beta_{h'a}^{l}\}} \braket{\{\beta_{ha}^{l}\} | \{\beta_{ga}^{l}\}}\\
\nn
&& - \ 2i\pi \delta(p_{i'}^0 - p_{i}^0) \braket{p_{i'}, \{\beta_{h'a}^{l}\} | M_e | p_i, \{\beta_{ha}^{l}\}} \delta^{(3)}(p_{i} - p_{i'}) \delta^{(3)}(p_{r'} - p_{r}) \delta^{(3)}(p_{r} - p_{r'}) \braket{\{\beta_{ia}^{l}\} | \{\beta_{i'a}^{l}\}} \braket{\{\beta_{r'a}^{l}\} | \{\beta_{ra}^{l}\}} \quad\\
&& \braket{\{\beta_{ga}^{l}\} | \{\beta_{g'a}^{l}\}} \braket{\{\beta_{i'a}^{l}\} | \{\beta_{r'a}^{l}\}} \braket{\{\beta_{ra}^{l}\} | \{\beta_{ia}^{l}\}} \braket{\{\beta_{g'a}^{l}\} | \{\beta_{h'a}^{l}\}} \braket{\{\beta_{ha}^{l}\} | \{\beta_{ga}^{l}\}}.
\ea
\ba
\nn
&& \braket{p_i, \{\beta_{ia}^{l}\} | S^{\dagger} | p_{i'}, \{\beta_{i'a}^{l}\}} \braket{\{\beta_{i'a}^{l}\} | \{\beta_{r'a}^{l}\}} \braket{p_{r'}, \{\beta_{r'a}^{l}\} | S | p_i, \{\beta_{fa}^{l}\}} \braket{\{\beta_{fa}^{l}\} | \{\beta_{ga}^{l}\}} \braket{p_g, \{\beta_{ga}^{l}\} | S^{\dagger} | p_{r'}, \{\beta_{g'a}^{l}\}}\\
\nn
&& \braket{\{\beta_{g'a}^{l}\} | \{\beta_{h'a}^{l}\}} \braket{p_{i'}, \{\beta_{h'a}^{l}\} | S | p_g, \{\beta_{la}^{l}\}} \braket{\{\beta_{la}^{l}\} | \{\beta_{ia}^{l}\}}\\
\nn
&& \simeq \delta^{(3)}(p_{i} - p_{i'}) \delta^{(3)}(p_{r'} - p_{i}) \delta^{(3)}(p_{g} - p_{r'}) \delta^{(3)}(p_{i'} - p_{g}) \braket{\{\beta_{ia}^{l}\} | \{\beta_{i'a}^{l}\}} \braket{\{\beta_{r'a}^{l}\} | \{\beta_{fa}^{l}\}} \braket{\{\beta_{ga}^{l}\} | \{\beta_{g'a}^{l}\}} \braket{\{\beta_{h'a}^{l}\} | \{\beta_{la}^{l}\}}\\
\nn
&& \braket{\{\beta_{i'a}^{l}\} | \{\beta_{r'a}^{l}\}} \braket{\{\beta_{fa}^{l}\} | \{\beta_{ga}^{l}\}} \braket{\{\beta_{g'a}^{l}\} | \{\beta_{h'a}^{l}\}} \braket{\{\beta_{la}^{l}\} | \{\beta_{ia}^{l}\}}\\
\nn
&& + \ 2i\pi \delta(p_{i}^0 - p_{i'}^0) \braket{p_{i}, \{\beta_{ia}^{l}\} | M_e^{\dagger} | p_{i'}, \{\beta_{i'a}^{l}\}} \delta^{(3)}(p_{r'} - p_{i}) \delta^{(3)}(p_{g} - p_{r'}) \delta^{(3)}(p_{i'} - p_{g}) \braket{\{\beta_{r'a}^{l}\} | \{\beta_{fa}^{l}\}} \braket{\{\beta_{ga}^{l}\} | \{\beta_{g'a}^{l}\}}\\
\nn
&& \braket{\{\beta_{h'a}^{l}\} | \{\beta_{la}^{l}\}} \braket{\{\beta_{i'a}^{l}\} | \{\beta_{r'a}^{l}\}} \braket{\{\beta_{fa}^{l}\} | \{\beta_{ga}^{l}\}} \braket{\{\beta_{g'a}^{l}\} | \{\beta_{h'a}^{l}\}} \braket{\{\beta_{la}^{l}\} | \{\beta_{ia}^{l}\}}\\
\nn
&& - \ 2i\pi \delta(p_{r'}^0 - p_{i}^0) \braket{p_{r'}, \{\beta_{r'a}^{l}\} | M_e | p_{i}, \{\beta_{fa}^{l}\}} \delta^{(3)}(p_{i} - p_{i'}) \delta^{(3)}(p_{g} - p_{r'}) \delta^{(3)}(p_{i'} - p_{g}) \braket{\{\beta_{ia}^{l}\} | \{\beta_{i'a}^{l}\}} \braket{\{\beta_{ga}^{l}\} | \{\beta_{g'a}^{l}\}}\\
\nn
&& \braket{\{\beta_{h'a}^{l}\} | \{\beta_{la}^{l}\}} \braket{\{\beta_{i'a}^{l}\} | \{\beta_{r'a}^{l}\}} \braket{\{\beta_{fa}^{l}\} | \{\beta_{ga}^{l}\}} \braket{\{\beta_{g'a}^{l}\} | \{\beta_{h'a}^{l}\}} \braket{\{\beta_{la}^{l}\} | \{\beta_{ia}^{l}\}}\\
\nn
&& + \ 2i\pi \delta(p_{g}^0 - p_{r'}^0) \braket{p_g, \{\beta_{ga}^{l}\} | M_e^{\dagger} | p_{r'}, \{\beta_{g'a}^{l}\}} \delta^{(3)}(p_{i} - p_{i'}) \delta^{(3)}(p_{r'} - p_{i}) \delta^{(3)}(p_{i'} - p_{g}) \braket{\{\beta_{ia}^{l}\} | \{\beta_{i'a}^{l}\}} \braket{\{\beta_{r'a}^{l}\} | \{\beta_{fa}^{l}\}}\\
\nn
&& \braket{\{\beta_{h'a}^{l}\} | \{\beta_{la}^{l}\}} \braket{\{\beta_{i'a}^{l}\} | \{\beta_{r'a}^{l}\}} \braket{\{\beta_{fa}^{l}\} | \{\beta_{ga}^{l}\}} \braket{\{\beta_{g'a}^{l}\} | \{\beta_{h'a}^{l}\}} \braket{\{\beta_{la}^{l}\} | \{\beta_{ia}^{l}\}}\\
\nn
&& - \ 2i\pi \delta(p_{i'}^0 - p_{g}^0) \braket{p_{i'}, \{\beta_{h'a}^{l}\} | M_e | p_g, \{\beta_{la}^{l}\}} \delta^{(3)}(p_{i} - p_{i'}) \delta^{(3)}(p_{r'} - p_{i}) \delta^{(3)}(p_{g} - p_{r'}) \braket{\{\beta_{ia}^{l}\} | \{\beta_{i'a}^{l}\}} \braket{\{\beta_{r'a}^{l}\} | \{\beta_{fa}^{l}\}}\\
&& \braket{\{\beta_{ga}^{l}\} | \{\beta_{g'a}^{l}\}} \braket{\{\beta_{i'a}^{l}\} | \{\beta_{r'a}^{l}\}} \braket{\{\beta_{fa}^{l}\} | \{\beta_{ga}^{l}\}} \braket{\{\beta_{g'a}^{l}\} | \{\beta_{h'a}^{l}\}} \braket{\{\beta_{la}^{l}\} | \{\beta_{ia}^{l}\}}\,.
\ea

Neglecting sub-subdominant interaction terms while focusing only on the zero-th order contributions, we may derive the tripartite mutual information for the asymptotically free particles, which finally reads 
\ba
\nn
&& I_{3(2)}^0 = S_2(C) + S_2(D) - S_2(AC) - S_2(AD)\\
\nn
&& = \log d_{A} + \log d_{B} + \log \Big[ \frac{1}{D^2} \Big(\frac{1}{\pi}\Big)^{2N} \int d^3p_i d^3p_r \delta^{(3)}(0) \delta^{(3)}(0) \int d^2\{\beta_{ia}^{l}\} d^2\{\beta_{ga}^{l}\} \braket{\{\beta_{ia}^{l}\} | \{\beta_{ia}^{l}\}} \braket{\{\beta_{ga}^{l}\} | \{\beta_{ga}^{l}\}} \Big]\\
\nn
&& + \log \Big[ \frac{1}{D^2} \Big(\frac{1}{\pi}\Big)^{N} \int d^3p_i d^3p_g \delta^{(3)}(p_{i} - p_{g}) \delta^{(3)}(p_{g} - p_{i}) \int d^2\{\beta_{ia}^{l}\} \braket{\{\beta_{ia}^{l}\} | \{\beta_{ia}^{l}\}} \Big]\\
&& = \log d_{A} + \log d_{B} + \log \Big[ \frac{d_A^2 d_B^2}{D^2} \Big] + \log \Big[ \frac{d_A d_B}{D^2} \Big] = 0.
\ea
This approximated result  actually conforms to na\"ive expectations that the free particles should not be scrambling.

We can easily revert this exemplified paradigm, taking interaction terms. 
Using Eq.~\eqref{D12} -- see e.g. Appendix \ref{A-D} ---  contributions that are provided to $S_2(AC)$ at leading-order by the interaction terms finally cast
\ba\label{trilo}
\nn
&& -\log\Big[ \frac{d_A^2 d_B^2}{D^2} + 2T \ \frac{d_A d_B}{D^2} \Big(\frac{1}{\pi}\Big)^{N} \int d^3p_i \int d^2\{\beta_{ia}^{l}\} \ \Big[ i \braket{p_i, \{\beta_{ia}^{l}\} | M_e^{\dagger} | p_{i}, \{\beta_{ia}^{l}\}} - i \braket{p_{i}, \{\beta_{ia}^{l}\} | M_e | p_i, \{\beta_{ia}^{l}\}} \Big] \Big]\\
\nn
&& = -\log\Big[ \frac{d_A^2 d_B^2}{D^2} - 4\pi T \ \frac{d_A d_B}{D^2} \Big(\frac{1}{\pi}\Big)^{2N} \int d^3p_{i} d^3p_{i'} \int d^2\{\beta_{ia}^{l}\} d^2\{\beta_{i'a}^{l}\} \delta(p_{i'}^0 - p_i^0) \big( 1+ 2\alpha B (p_i, p_{i'}) + 2\alpha \tilde{B} (p_i, p_{i'}) + o(\alpha) \big)\\
\nn
&& \exp \Big\{ - \sum_{l,a} |\epsilon_{i'a}^{l} -  \epsilon_{ia}^{l}|^2 \Big\} \Big\{ |m_{0,0} (p_{i}, p_{i'})|^2 + o(\alpha^2) \Big\} \Big]\\
\nn
&& = -\log\Big[ \frac{d_A^2 d_B^2}{D^2} - 4\pi T \ \frac{d_A d_B}{D^2} \Big(\frac{1}{\pi}\Big)^{2N} \int d^3p_{i} d^3p_{i'} \int d^2\{\beta_{ia}^{l}\} d^2\{\beta_{i'a}^{l}\} \delta(p_{i'}^0 - p_i^0) \big( 1+ 2\alpha B (p_i, p_{i'}) + 2\alpha \tilde{B} (p_i, p_{i'}) + o(\alpha) \big)\\
\nn
&& \exp \Big\{ - \sum_{l,a} |\beta_{i'a}^{l} - \beta_{a}^{l}(p_{i'}) -  \beta_{ia}^{l} + \beta_{a}^{l}(p_{i})|^2 \Big\} \Big\{ |m_{0,0} (p_{i}, p_{i'})|^2 + o(\alpha^2) \Big\} \Big]\\
\nn
&& \simeq -\log\Big[ \frac{d_A^2 d_B^2}{D^2} - 4\pi T \ \frac{d_A d_B}{D^2} \Big(\frac{1}{\pi}\Big)^{2N} \int d^3p_{i} d^3p_{i'} \int d^2\{\beta_{ia}^{l}\} d^2\{\beta_{i'a}^{l}\} \delta(p_{i'}^0 - p_i^0) \big( 1+ 2\alpha B (p_i, p_{i'}) + 2\alpha \tilde{B} (p_i, p_{i'}) + o(\alpha) \big)\\
\nn
&& \exp \Big\{ - \sum_{l,a} |\beta_{i'a}^{l}  -  \beta_{ia}^{l}|^2 \Big\} \Big\{ |m_{0,0} (p_{i}, p_{i'})|^2 + o(\alpha^2) \Big\} \Big]\\
\nn
&& = -\log\Big[ \frac{d_A^2 d_B^2}{D^2} - 4\pi \frac{d_A d_B^2}{D^2} T \int d^3p_{i} d^3p_{i'} \delta(p_{i'}^0 - p_i^0) \big( 1+ 2\alpha B (p_i, p_{i'}) + 2\alpha \tilde{B} (p_i, p_{i'}) + o(\alpha) \big) \Big\{ |m_{0,0} (p_{i}, p_{i'})|^2 + o(\alpha^2) \Big\} \Big]\\
\nn
&& = -\log\Big[ \frac{d_A^2 d_B^2}{D^2} - 4\pi \frac{d_A d_B^2}{D^2} T \int d^3p_{i} d^3p_{i'} \delta(p_{i'}^0 - p_i^0) \Big\{ |m_{0,0} (p_{i}, p_{i'})|^2 + o(\alpha^2) \Big\} \Big]\\
&& = - \log \Big[ \frac{d_A^2 d_B^2}{D^2} \Big] - \log \Big[ 1 - \frac{4 \pi T}{d_A} \int d^3p_i d^3p_{i'} \ \delta(p_{i'}^0-p_i^0) |m_{0,0} (p_{i}, p_{i'})|^2 \Big]\,.
\ea
\\

The complex coefficients $\beta^l_a(p_i)$ and $\beta^l_a(p_{i'})$ that appears in the second hand-side or Eq.~\eqref{trilo} both satisfy the convergence conditions $\sum_{l,a} |\beta^l_a(p_i)|^2 < \infty$ and $\sum_{l,a} |\beta^l_a(p_{i'})|^2 < \infty$. Thus both the coefficients $\beta^l_a(p_i)$ and $\beta^l_a(p_{i'})$ are finite and can be regarded as finite displacements not affecting the result of the integration. The integrand $m_{0,0} (p_{i}, p_{i'})$ appearing in the latter hand-side of Eq.~\eqref{trilo} individuates the first-order-expansion of the $M$ matrix $\braket{p_{i'}, \{\beta_{i'a}^{l}\} | M_e | p_{i}, \{\beta_{ia}^{l}\}}$. The expression of this latter casts, for $|p_i|=|p_{i'}|$ \cite{QFTMS}
\be
|m_{0,0} (p_{i}, p_{i'})|^2 = \frac{(2 \pi \alpha Z)^2}{4 m^2 |p_i|^4 \sin^4(\theta_i/2)} \big[ |p_i|^2 \cos^2(\theta_i/2) + m^2 \big]\,.
\ee
Similarly, the contribution of the interaction terms to $S_2(AD)$ at the leading-order provides the result
\ba
\nn
&& -\log\Big[ \frac{d_A d_B}{D^2} + \frac{2T}{D^2} \Big(\frac{1}{\pi}\Big)^{N} \int d^3p_{i} \int d^2\{\beta_{ia}^{l}\} \Big[ i \braket{p_{i}, \{\beta_{ia}^{l}\} | M_e^{\dagger} | p_{i}, \{\beta_{ia}^{l}\}} - i \braket{p_{i}, \{\beta_{ia}^{l}\} | M_e | p_{i}, \{\beta_{ia}^{l}\}} \Big] \Big]\\
&& = - \log \Big[ \frac{d_A d_B}{D^2} \Big] - \log \Big[ 1 - \frac{4 \pi T}{d_A} \int d^3p_i d^3p_{i'} \ \delta(p_{i'}^0-p_i^0) |m_{0,0} (p_{i}, p_{i'})|^2 \Big]\,.
\ea
\end{widetext}
Thus, the contribution of the interaction terms to the evaluation of $I_{3(2)}$ at the leading order are expressed by\\
\ba
\nn
I_{3(2)} &=& S_2(C) + S_2(D) - S_2(AC) - S_2(AD)\\
\nn
&=& \log d_{A} + \log d_{B} + \log \Big[ \frac{d_A^2 d_B^2}{D^2} \Big] +\log \Big[ \frac{d_A d_B}{D^2} \Big]
\nn \\
&+& 2\log \Big[ 1 
\!-\! \frac{4 \pi T}{d_A} \!\int \! d^3p_i d^3p_j  \delta(p_j^0-p_i^0) |m_{0,0} (p_{i}, p_{j})|^2 \Big] \nn \\
&=& 2 \log \Big[ 1 \!- \!\frac{4 \pi T}{d_A} \!\int \! d^3p_i d^3p_j  \delta(p_j^0-p_i^0) |m_{0,0} (p_{i}, p_{j})|^2 \Big] \nn\\
&\simeq& - \frac{8 \pi T}{d_A} \int d^3p_i d^3p_j  \delta(p_j^0-p_i^0) |m_{0,0} (p_{i}, p_{j})|^2 \nn\\
&=& - \frac{8 \pi T}{d_A} \int d|p_i| d\theta_i \ 16\pi^2 \cos(\theta_i/2) \nn \\
&\phantom{=}& \times \frac{(2 \pi \alpha Z)^2}{4 m^2 \sin^3(\theta_i/2)} \big[ |p_i|^2 \cos^2(\theta_i/2) + m^2 \big]\,,
\ea
having used that $D=d_A d_B$.\\

The final result can be the achieved specifying the integral with respect to $\theta_i$ and the integral with respect to $p_i$. Even though the integral with respect to $\theta_i$ diverges at $\theta_i=0$, we shall not be concerned by such a non physical divergence. Guidance in stead is provided by the  experimental results, which display a peak at $\theta_i=0$ that cannot be interpreted as a divergence. \\

Expanding \eqref{pf}, we can isolate two terms, and use $A_1$ and $A_2$ to label the two integrals with respect to $\theta_i$, which are positive for $\theta_i \in [0,\pi]$. Then the tripartite mutual information $I_{3(2)}$ casts
\ba
I_{3(2)} &=& - \frac{8 \pi T}{d_A} \ \frac{4\pi^2 (2 \pi \alpha Z)^2}{m^2} \ A_1 \int |p_i|^2 d|p_i| \nn\\
&\phantom{=}& - \frac{8 \pi T}{d_A} \ 4\pi^2 (2 \pi \alpha Z)^2 \ A_2 \int d|p_i|\,.
\ea

To calculate the integral with respect to $p_i$ some additional steps are required. Considering that inside a box of volume $V$ momenta are discretized as ${\bf p}= {\bf n} (2\pi/L)$, where $V=L^3$ and ${\bf n}=(n_1, n_2, n_3)$, with $n_1, n_2, n_3\in \mathbb{N}$, we can express the dimension $d_A$ of the space of hard particle states $A$ as the sum over the modes ${\bf n}$, namely    
\ba
 \int d^3p = 
\frac{(2\pi)^3}{V} \, \sum_{n} = (2\pi)^3 \frac{d_A}{V}\,,
\ea
and then immediately find a useful related equality   
\ba
\int d|p| = |p| = 2\pi \left( \frac{3 d_A}{4\pi V} \right)^{\frac{1}{3}}.
\ea

We are ready to combine all the contributions, and finally obtain as the tripartite mutual information $I_{3(2)}$, which reads
\ba
I_{3(2)} &=& - 8\pi^2 (2 \pi \alpha Z)^2 A_1 (2\pi)^3 \frac{T}{m^2 V} \nn\\
&\phantom{a}& - 64\pi^4 (2 \pi \alpha Z)^2 A_2 \big( \frac{3}{4\pi}\big)^{\frac{1}{3}} d_A^{-\frac{2}{3}} \frac{T}{V^{\frac{1}{3}}}\,.
\ea

\bibliography{references}

\end{document}